\documentclass[twocolumn]{aastex631}

\usepackage{amsmath}

\definecolor{fuchsia}{rgb}{0.54, 0.17, 0.89}
\definecolor{azure}{rgb}{0.0, 0.5, 1.0}
\definecolor{pgreen}{rgb}{0.12, 0.3, 0.17}
\definecolor{alizarin}{rgb}{0.82, 0.1, 0.26}

\newcommand{\kms}{{\rm km~s^{-1}}}
\newcommand{\oii}{[\textrm{O}~\textsc{ii}]}
\newcommand{\oiii}{[\textrm{O}~\textsc{iii}]}
\newcommand{\oiiir}{[\textrm{O}~\textsc{iii}]_{\lambda5007}}

\newcommand{\oiiif}{$[\textrm{O}~\textsc{iii}]_{\lambda4363}$}

\newcommand{\simgt}{\,\rlap{\lower 3.5 pt \hbox{$\mathchar \sim$}} \raise
1pt \hbox {$>$}\,}
\newcommand{\simlt}{\,\rlap{\lower 3.5 pt \hbox{$\mathchar \sim$}} \raise
1pt \hbox {$<$}\,}
\newcommand{\Msun}{M_{\odot}}

\newcommand{\logm}{\log M_*/\Msun}

\newcommand{\logoh}{$\log {\rm (O/H)}$}

\newcommand{\ly}{${\rm Ly\alpha}$}

\newcommand{\hb}{${\rm H\beta}$}

\newcommand{\Z}{{\rm [Z/H]}}

\newcommand{\id}{A2744-z7p9}

\submitjournal{ApJ}

\shorttitle{Metallicity Scatter in the Core of a Protocluster at $z=7.88$}
\shortauthors{Morishita, Stiavelli, et al.}

\graphicspath{{./}{figures/}}

\begin{document}

\title{Metallicity Scatter Originating from Sub-kiloparsec Starbursting Clumps in the Core of a Protocluster at z=7.88}

\correspondingauthor{Takahiro Morishita}
\email{takahiro@ipac.caltech.edu}

\author[0000-0002-8512-1404]{Takahiro Morishita}
\affiliation{IPAC, California Institute of Technology, MC 314-6, 1200 E. California Boulevard, Pasadena, CA 91125, USA}

\author[0000-0001-9935-6047]{Massimo Stiavelli}
\affiliation{Space Telescope Science Institute, 3700 San Martin Drive, Baltimore, MD 21218, USA}
\affiliation{The William H. Miller III, Dept. of Physics \& Astronomy, Johns Hopkins University, Baltimore, MD 21218, USA}
\affiliation{Dept. of Astronomy, University of Maryland, College Park, MD 20742, USA}

\author[0000-0002-5057-135X]{Eros~Vanzella}
\affiliation{INAF -- OAS, Osservatorio di Astrofisica e Scienza dello Spazio di Bologna, via Gobetti 93/3, I-40129 Bologna, Italy}

\author[0000-0003-1383-9414]{Pietro~Bergamini}
\affiliation{Dipartimento di Fisica, Università degli Studi di Milano, Via Celoria 16, I-20133 Milano, Italy}
\affiliation{INAF -- OAS, Osservatorio di Astrofisica e Scienza dello Spazio di Bologna, via Gobetti 93/3, I-40129 Bologna, Italy}

\author[0000-0003-4109-304X]{Kristan Boyett}
\affiliation{Department of Physics, University of Oxford, Denys Wilkinson Building, Keble Road, Oxford OX1 3RH, UK}

\author[0000-0003-1564-3802]{Marco Chiaberge}
\affiliation{Space Telescope Science Institute for the European Space Agency (ESA), ESA Office, 3700 San Martin Drive, Baltimore, MD 21218, USA}
\affiliation{The William H. Miller III Department of Physics and Astronomy, Johns Hopkins University, Baltimore, MD 21218, USA}

\author[0000-0002-5926-7143]{Claudio Grillo}
\affiliation{Dipartimento di Fisica, Università degli Studi di Milano, Via Celoria 16, I-20133 Milano, Italy}
\affiliation{INAF - IASF Milano, via A. Corti 12, I-20133 Milano, Italy}

\author[0000-0003-4570-3159]{Nicha Leethochawalit}
\affiliation{National Astronomical Research Institute of Thailand (NARIT), Mae Rim, Chiang Mai, 50180, Thailand}

\author[0000-0003-1427-2456]{Matteo Messa}
\affiliation{INAF -- OAS, Osservatorio di Astrofisica e Scienza dello Spazio di Bologna, via Gobetti 93/3, I-40129 Bologna, Italy}

\author[0000-0002-4140-1367]{Guido Roberts-Borsani}
\affiliation{Department of Astronomy, University of Geneva, Chemin Pegasi 51, 1290 Versoix, Switzerland}
\affiliation{Department of Physics \& Astronomy, University College London, London, WC1E 6BT, UK}

\author[0000-0002-6813-0632]{Piero Rosati}
\affiliation{INAF -- OAS, Osservatorio di Astrofisica e Scienza dello Spazio di Bologna, via Gobetti 93/3, I-40129 Bologna, Italy}
\affiliation{Dipartimento di Fisica e Scienze della Terra, Università degli Studi di Ferrara, Via Saragat 1, I-44122 Ferrara, Italy}

\author[0000-0002-5558-888X]{Anowar J. Shajib}
\affiliation{Department of Astronomy and Astrophysics, University of Chicago, Chicago, IL 60637, USA}
\affiliation{Kavli Institute for Cosmological Physics, University of Chicago, Chicago, IL 60637, USA}
\affiliation{NHFP Einstein Fellow}



\begin{abstract}
We present new JWST NIRSpec integral field unit (IFU) G395H/F290LP observations of a merging galaxy system at $z=7.88$, part of A2744-z7p9, the most distant protocluster to date. The IFU cube reveals $\oiiir$ emissions in two previously known galaxies (ZD3 and ZD6) and a newly identified galaxy, ZD12, at $z_{\rm spec}=7.8762$. One of the detected \oiii-emitting regions has a detection of the auroral \oiiif\, line, allowing us to derive a direct metallicity of \logoh\,$+12=7.4\pm0.2$, while metallicities in other regions are measured using strong line calibration methods. We find large deviations within the measured metallicity ($\Delta \log {\rm (O/H)}\sim1$), which suggests a fast chemical enrichment from intense star formation and merger-driven growth, as expected in early galaxies. Our analysis shows that metal-poor regions could easily be outshone by more enriched regions, posing a challenge for spectroscopic analysis based on integrated light (i.e., NIRSpec MSA) against identifying metal-free star formation in the early universe. NIRCam imaging reveals seven UV-bright clumps in ZD12, in the range of stellar mass $\logm\sim7.6$--8.9. Four of them are unresolved ($\simlt 100$\,pc) and intensely star-forming ($>30\,M_\odot\, {\rm yr^{-1}\, kpc^{-2}}$), likely contributing to the scatter in metallicity by producing an ideal environment for rapid chemical cycles. Lastly, we revisit the nature of the host protocluster by including new member galaxies identified here and in the literature, and obtain local overdensity factor $\delta=44_{-31}^{+89}$, total halo mass $M_{\rm h} = 5.8_{-0.3}^{+0.2}\times10^{11}\,M_\odot$, and a formal velocity dispersion of $1100\pm500$\,km\,s$^{-1}$.
\end{abstract}

\keywords{}


\section{Introduction} \label{sec:intro}

The gas cycle within a galaxy system plays an important role in shaping the host structure and establishing its observed properties. The ejecta from previously formed stars enrich surrounding gas, some of which contribute to subsequent star formation cycles, and enhance chemical abundances within the system. Galaxies are by no means closed systems. A series of interplays with other galaxies and the cosmic web through the accretion of pristine gas adds further complications in establishing the chemical enrichment journey over the lifetime of a galaxy \citep[e.g.,][]{oppenheimer06,tumlinson17,chisholm18}. 

Despite such complexities, the accumulation of stars evolved in such a cyclical way represents a major component of galaxies in a later epoch, culminating with an established relation of metal contents and other fundamental galaxy properties, such as stellar mass, i.e., the mass--metallicity relation, seen in the local universe \citep[e.g.,][]{tremonti04,gallazzi05,andrews13,kirby13,curti20}. 
Studies, enabled by the JWST, have extended the investigation to higher redshifts. The correlation between the two parameters seems to exist beyond the local universe, to $z>3$ \citep{nakajima23,sanders23,laseter23,curti23,chemerynska24}, but with an offset to the local relation \citep[e.g.,][]{heintz23,morishita24}. Studying the relationship in earlier times gives us insight into how different physical mechanisms act on the formation of the tight correlation \citep[e.g.,][]{ma16,pallottini24}. 

In this regard, those early JWST observations of high-$z$ galaxies are already informative, finding considerably large scatters around the relation; at $z\sim0$ the scatter is found to be $\sim0.1$ \citep{curti20}, while the observed scatter is significantly increased to $\simgt0.3$\,dex at $z>5$. \citet{morishita24} investigated this in detail and discussed the possible origins. The study found that the scatter could not be reduced by the introduction of an additional parameter in the relationship, unlike the case for low-$z$ galaxies with star formation rate as the third parameter \citep[e.g.,][]{mannucci10}. Instead, the scatter was found to be related to systematics coming from the limited slit size of NIRSpec/MSA and the exact slit position within the target galaxy. In addition, the same study argues that, given that many MSA observations put the slit in the photocenter of galaxies by default, the measured metallicity with NIRSpec/MSA may be biased toward a more enriched region. However, this explanation is possible only when there are significant metallicity variations within individual galaxies, which is not obvious in early galaxies, thus requiring further investigation with spatially resolved spectroscopy. 

Recent studies using JWST/IFU have shown various conclusions; \citet{marconcini24} found small metallicity variations ($\Delta \log {\rm O/H} \simlt 0.2$\,dex) within a luminous galaxy system CR7 at $z=6.6$ \citep[also][for similar findings in COS-3018 at $z=6.85$ and B14-65666 at $z=7.152$, respectively]{scholtz24,jones24}. On the other hand, \citet{marconcini24a} reported larger deviations ($\simgt0.5$\,dex) within MACS1149-JD1 at $z=9.11$. The wealth of high-resolution data provided by the JWST now enables us to explore the physical mechanisms driving such deviations.

In this paper, we present a new $R\sim3000$ IFU observation executed in the Cycle~2 \citep[PID~4553;][]{stiavelli23}, toward a unique merging galaxy system at $z=7.88$. The system is located in the core of the most distant protocluster, A2744-z7p9, to date \citep{zheng14,laporte14,ishigaki16,roberts-borsani22,morishita23b,hashimoto23}. Furthermore, being behind the massive cluster, Abell~2744, the system is magnified by a factor of $\mu\sim2$, enabling us to study the inter-stellar medium (ISM) properties and sub-galactic structures in great detail. In Sec.~\ref{sec:data}, we present our data reduction and analyses. In Sec.~\ref{sec:ana}, we measure the physical properties of the identified member galaxies via emission line analysis and spectral energy distribution fitting. In Sec.~\ref{sec:disc}, we discuss the inferred properties of those galaxies in the context of the mass--metallicity relationship and provide updated characterizations of the host protocluster \id. Where relevant, we adopt the AB magnitude system \citep{oke83,fukugita96}, cosmological parameters of $\Omega_{\rm m}=0.3$, $\Omega_\Lambda=0.7$, $H_0=70\,\kms\, {\rm Mpc}^{-1}$, and the \citet{chabrier03} initial mass function (IMF). Throughout the manuscript, we use the term metallicity to refer to oxygen abundance.

\section{Data}\label{sec:data}

\subsection{NIRSpec/IFU Observations and Reduction}\label{sec:ifu}
Our NIRSpec/IFU observations (GTO4553, PI Stiavelli) were executed within a single visit on Oct 27, 2024, centered at (R.A., Decl.) = ($3.60680, -30.38082$). The visit was configured at the V3 position angle PA\_V3=41.1\,degree with the high-resolution grating G395H/F290LP. The exposure consists of 12 dithers, with the total science time 19,432\,sec (7.44\,hrs including overhead was charged). 

The data are reduced using the official JWST pipeline (ver 1.16.0) with additional custom steps. Briefly, we retrieve the level1 products (\_uncal.fits) from MAST. We run the stage1 step and then the stage2 step, with additional stripe elimination using NSClean \citep{rigby23}\footnote{\url{https://github.com/JWST-Templates/NSClean}} and cosmis ray rejection using {\tt lacosmic} \citep{vandokkum01,bradley23}. We also run the source detection tool {\tt SExtractor} \citep{bertin96} in each \_rate image and flag significant point-like source, i.e., cosmic ray (CR) that are not detected in the previous steps. We run the stage3 step to construct a cube and subtract background in each wavelength frame using off-source regions. We run {\tt lacosmic} again in each wavelength frame to mask any residual CRs.

\begin{figure*}
\centering
    \includegraphics[width=1\textwidth]{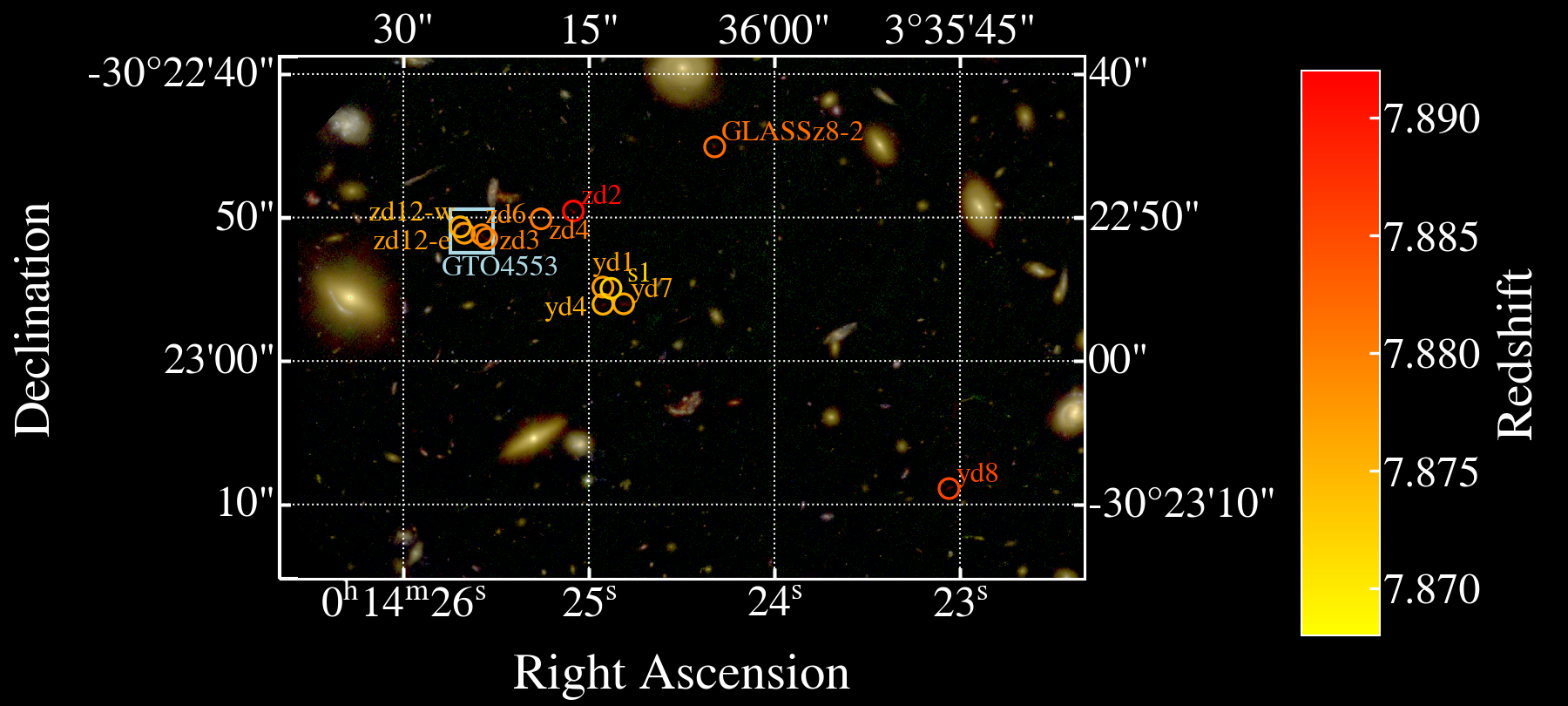}
    \includegraphics[width=1\textwidth]{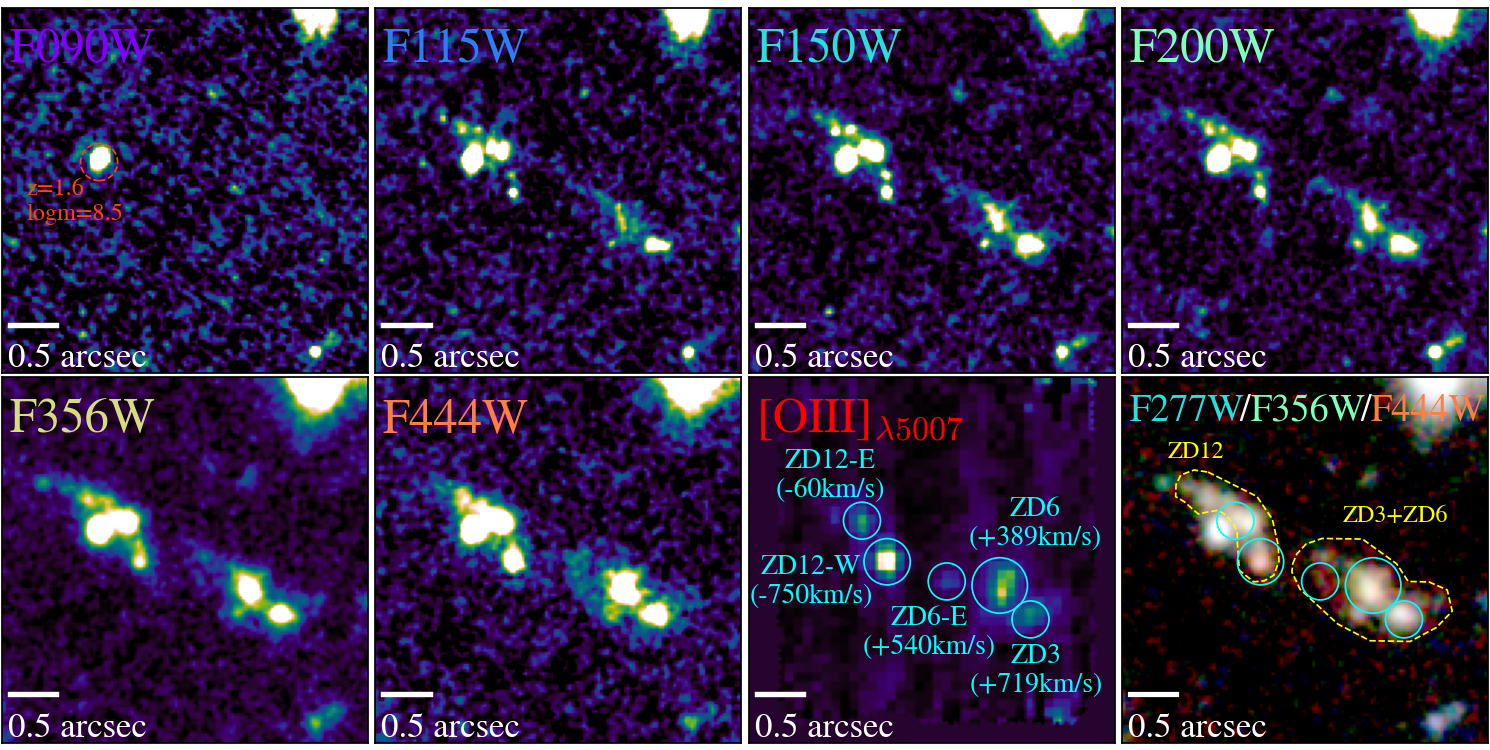}
	\caption{
    $Top$: Mosaic image of the field (blue: HST/ACS-F435W, green: NIRCam-F090W, red: NIRCam-F150W) centered on the protocluster, \id. The spectroscopically confirmed members are marked (circles). The Field-of-View of our IFU observations GTO4553 is shown (blue rectangles).
    $Bottom$: Postage stamps of NIRCam images ($4''\times4''$), showing the region of the GTO4553 observations. 
    Five $\oiii$-emitting regions analyzed in the main texts are marked in the $\oiiir$-line map (cyan circles) with the velocity offset measurement from the systemic redshift $z_{\rm sys}=7.8784$.
 }
\label{fig:stamp}
\end{figure*}

\subsection{Imaging Data and Photometry}\label{sec:data-jwst}
We utilize the JWST NIRCam imaging data available in the field. Those include; GLASS-JWST \citep[ERS1324;][]{treu22}, UNCOVER \citep[GO2561;][]{bezanson22}, DD2756 \citep{chen22}, GO2883 (PI F. Sun), All the Little Things \citep[ALT; GO3516;][]{naidu24}, GO3538 (PI E. Iani), BEACON \citep[GO3990;][]{morishita24beacon}, and Medium Bands, Mega Science \citep[GO4111;][]{suess24}. We also utilize the \textit{Hubble} Space Telescope (HST) Advanced Camera for Surveys (ACS) and Wide Field Camera 3 (WFC3)-IR mosaic images publicly available by the GLASS-JWST team \citep{merlin22}, which combined data from multiple HST programs \citep{postman12,lotz17,kelly18,steinhardt20}.

All data were compiled and used for photometry as described in \citet{morishita24c}. Briefly, we identified sources in the IR detection image (F277W+F356W+F444W stack) using \texttt{SExtractor} \citep{bertin96}, and measured fluxes on the mosaic images {(matched to the F444W point-spread function, PSF)}, with a fixed aperture of radius $r=0.\!''16$\,. The measured fluxes are calibrated using the method introduced in \citet{morishita23}. Interested readers are referred to the work for more details. The filters included in our analysis and the limiting magnitudes ($5\,\sigma$, for a point source) are: HST-F435W (28.7), F606W (28.1), F775W (28.3), F814W (27.7), F105W (27.5), F125W (27.5), F140W (28.2), F160W (27.1), NIRCam-F070W (28.5), F090W (28.9), F115W (28.7), F140M (27.9), F150W (28.8), F162M (27.9), F182M (28.1), F200W (28.9), F210M (28.0), F250M (28.0), F277W (29.1), F300M (28.1), F335M (28.1), F356W (29.1), F360M (28.2), F410M (28.6), F430M (27.7), F444W (28.9), F460M (26.8), F480M (27.2), NIRISS-F115W (28.8), F150W (28.7), F200W (28.6), F356W (28.8), F430M (27.6), F444W (28.5), F480M (27.0).

In the following analysis, we adopt the latest magnification model by \citet{bergamini23,bergamini23b}.

\section{Analysis and Results}\label{sec:ana}

\begin{figure*}
\centering
    \includegraphics[width=0.49\textwidth]{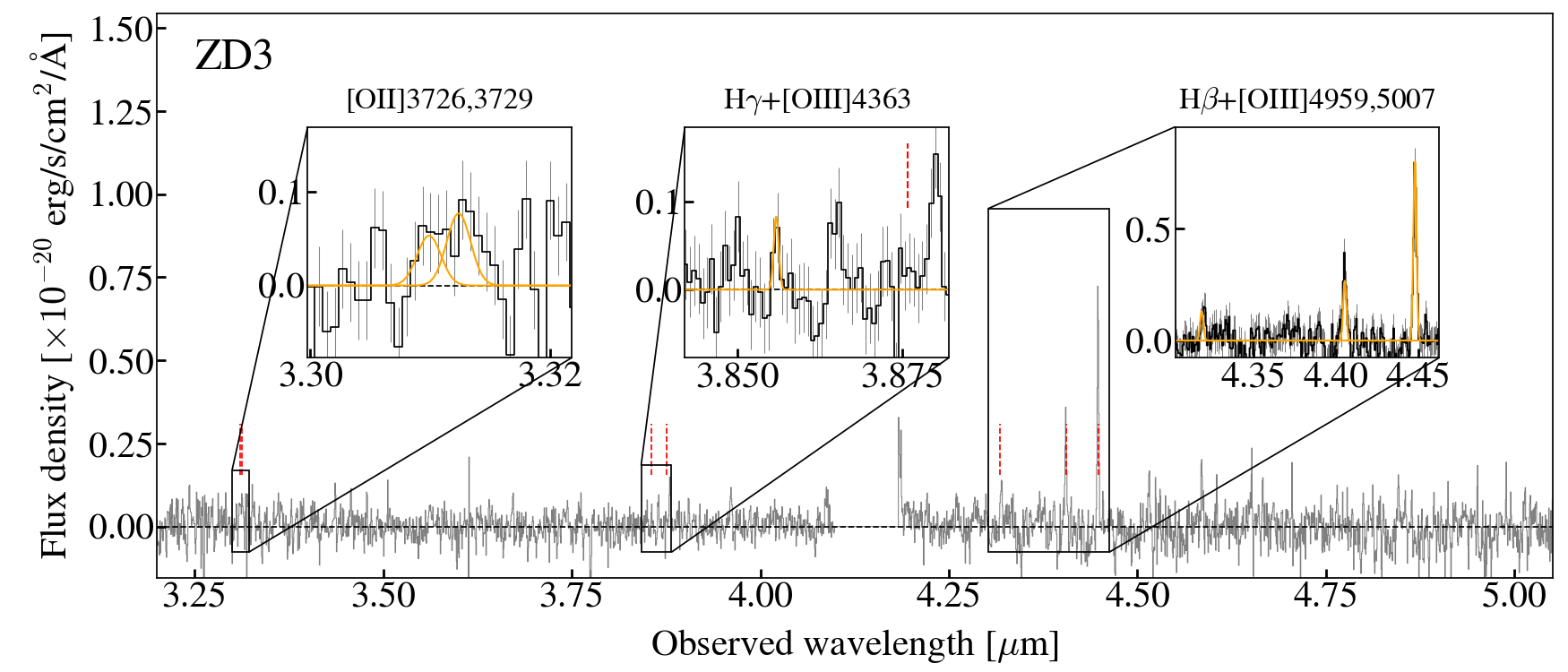}
    \includegraphics[width=0.49\textwidth]{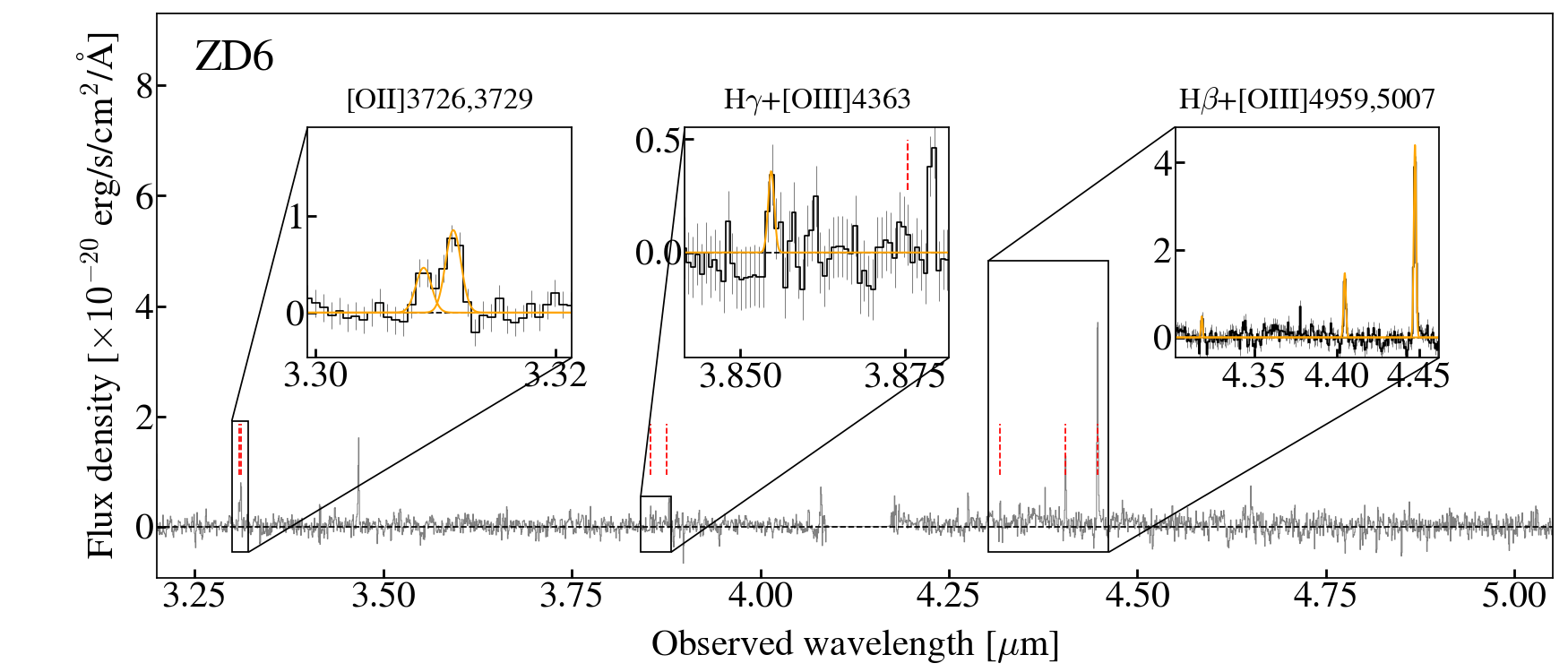}
    \includegraphics[width=0.49\textwidth]{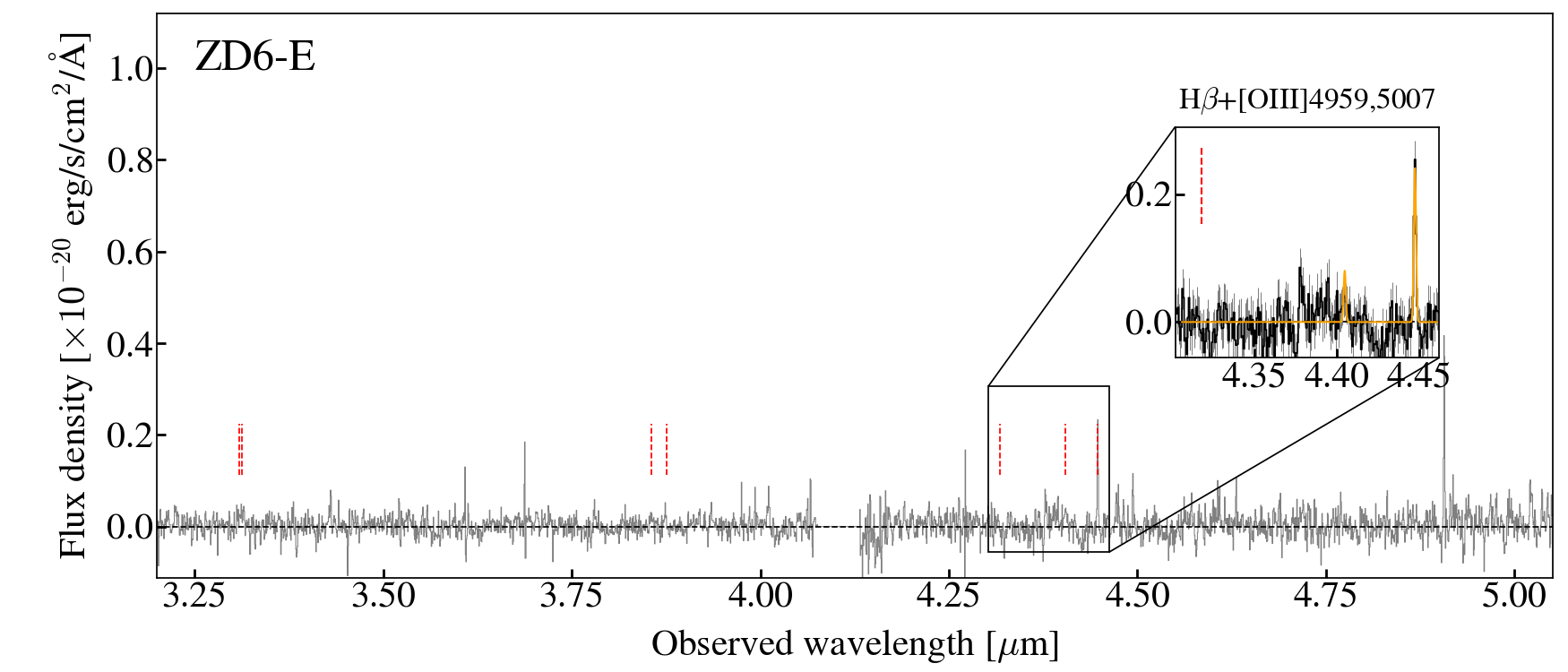}
    \includegraphics[width=0.49\textwidth]{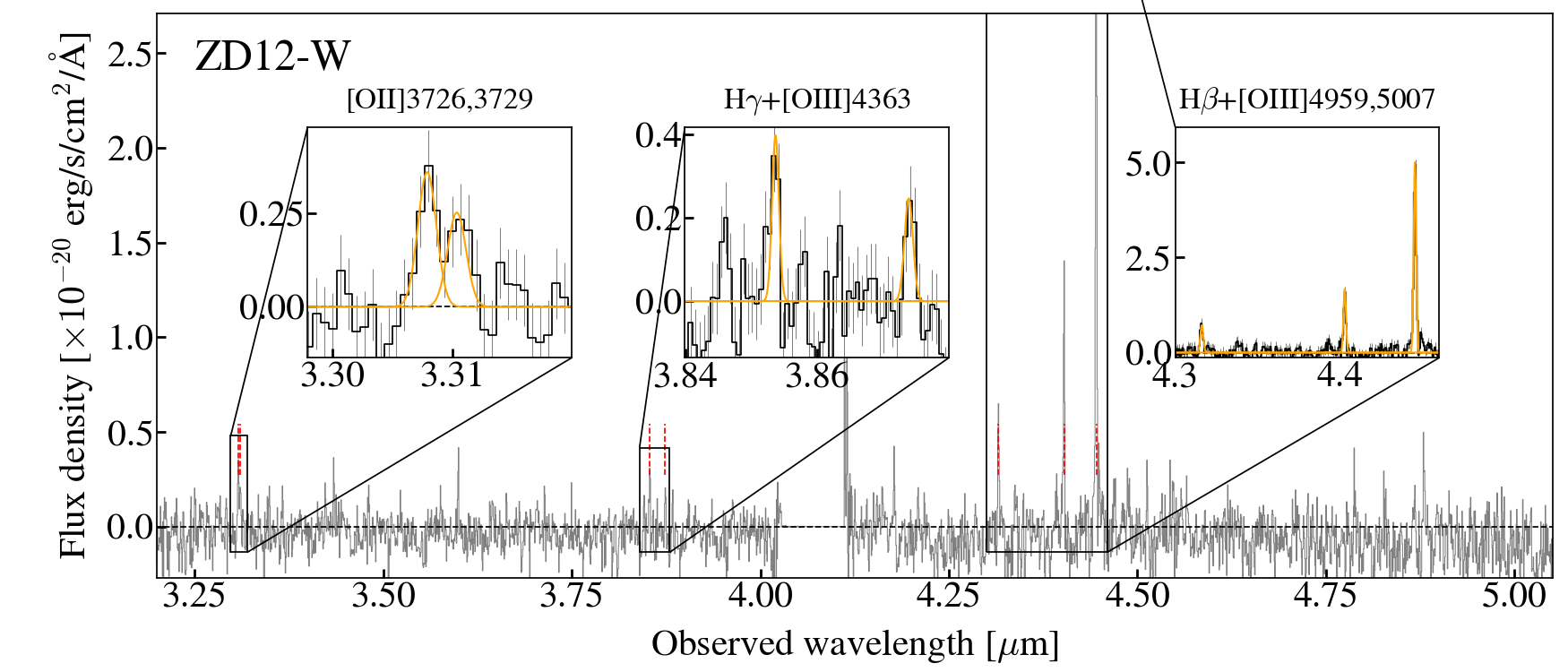}
    \includegraphics[width=0.49\textwidth]{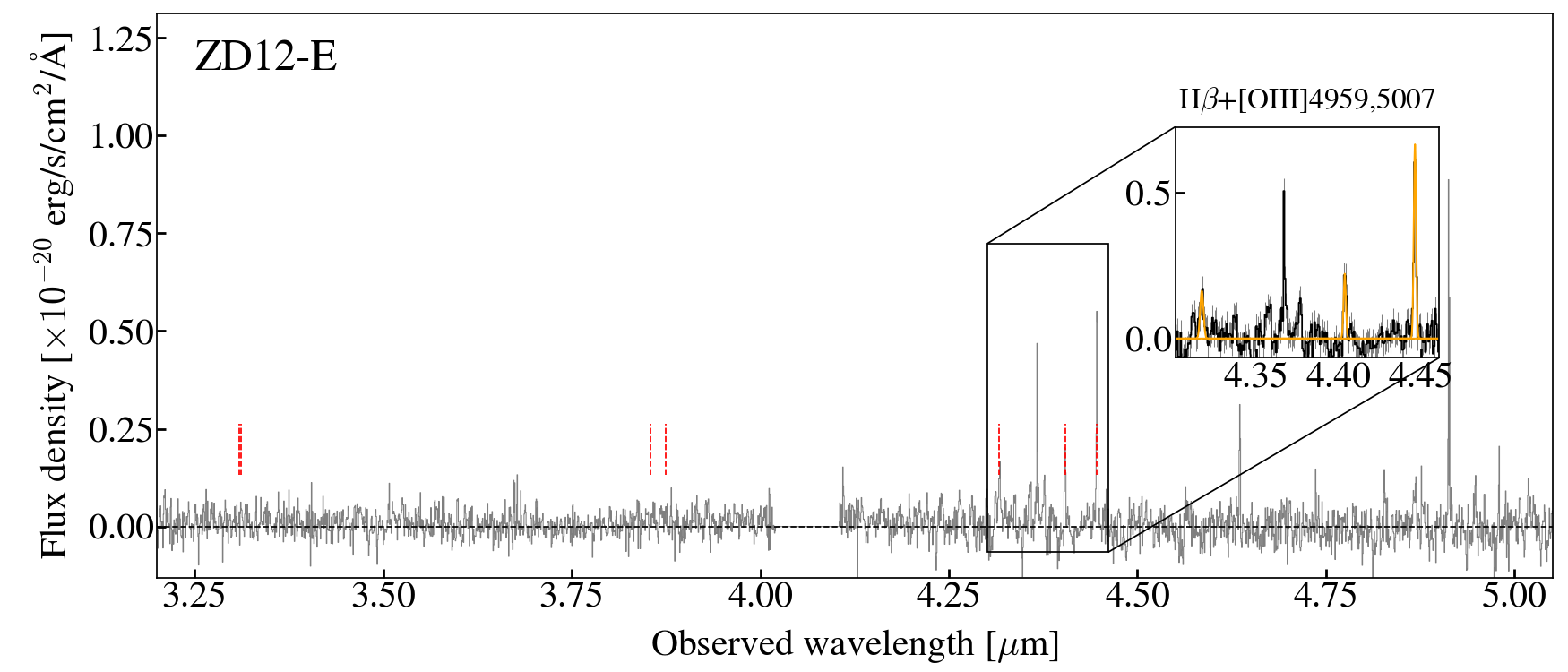}
    \includegraphics[width=0.49\textwidth]{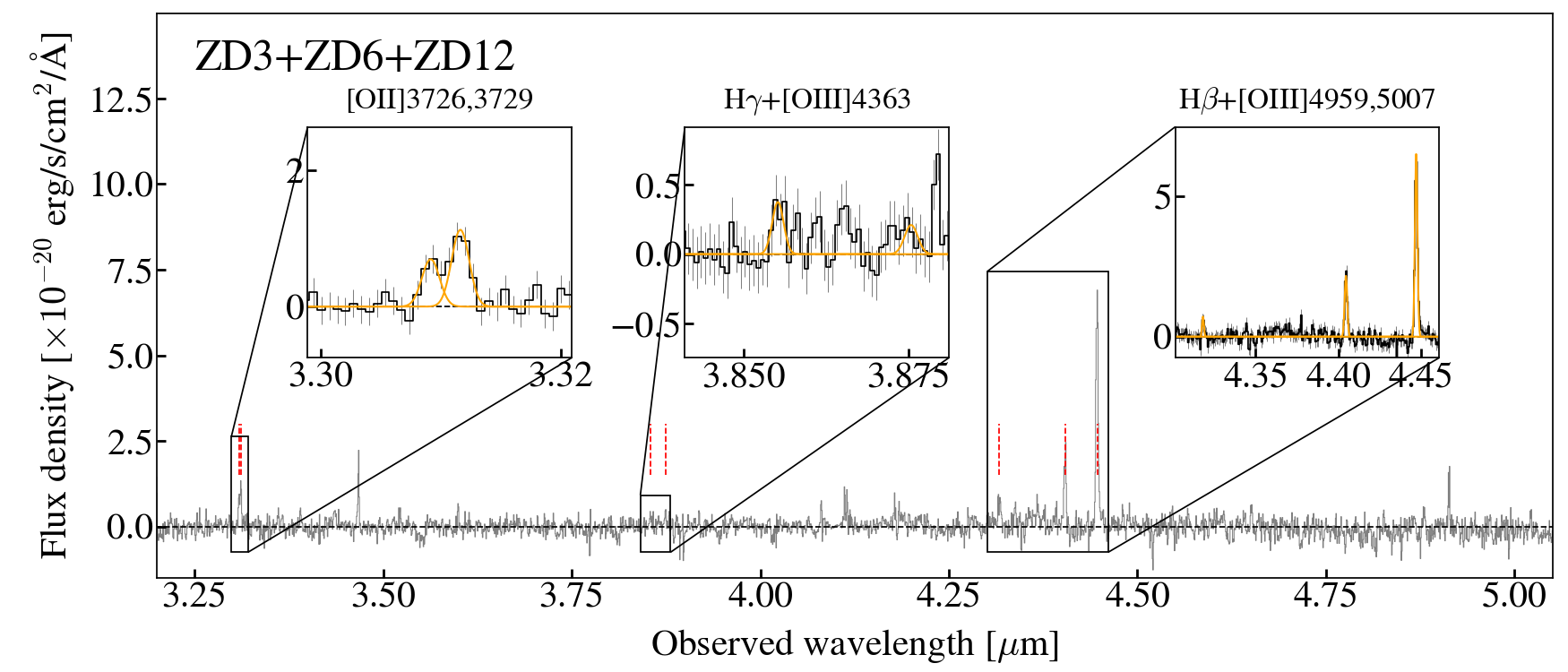}
	\caption{
    Extracted G395H/F290LP spectra of the \oiii-emitting regions (black lines). The defined regions for each spectrum are shown in Fig.~\ref{fig:stamp}. Detected emission lines (SNR\,$>1.5$) are shown in the inset, where the fitted Gaussian models (orange) are shown.
 }
\label{fig:spectra}
\end{figure*}

\subsection{Emission Line Analysis}\label{linefit}
Emission lines are extracted using the reduced flux cube. Given the limited Signal-to-Noise ratio (SNR), we mainly base our analysis on the strongest emission line, $\oiiir$. We extract the $\oiiir$ emission line map by coadding cube frames (each weighted by the corresponding rms frame) over the velocity range $\pm300$\,km\,s$^{-1}$ centered at $z=7.88$, i.e., the systemic redshift of ZD3 and ZD6. 

In Fig.~\ref{fig:stamp}, we show the extracted $\oiiir$ emission line map along with the NIRCam images. We identify five separate regions detected with SNR\,$>3$. Two are ZD3 and ZD6, previously spectroscopically confirmed galaxies \citep{morishita23b}. In addition, one emitting region is detected east of ZD6, hereafter referred to as ZD6-E. Two strong \oiii-emitting regions are identified within the eastern system ZD12\footnote{\citet{zheng14} presented 11 $z$-band dropout sources. ZD12 was not identified in the original photometric study, possibly due to confusion with the foreground galaxy.}, which we refer to as ZD12-W and ZD12-E, respectively. In what follows, we base our analysis on those five regions. We refer to them as \oiii-{\it emitting regions}, to distinguish them from UV {\it clumps} that are defined separately later in this paper.

Spectra are extracted within apertures of $\sim0.\!''15$--$0.\!''3$ centered in the \oiii-emitting region, where the aperture size is determined by a growth curve method. Furthermore, we define larger segments for the entire ZD3 + ZD6, ZD12, and ZD3 + ZD6 + ZD12 systems (regions refined by yellow lines in Fig.~\ref{fig:stamp}) and extract the spectra for further analysis. The extracted spectrum of the total system is shown in Fig.~\ref{fig:spectra}.  

We note a foreground galaxy (ID 51545, 00:14:25.7, $-$30:22:50.6), to the South-East of ZD12 (Fig.~\ref{fig:stamp}). This galaxy is clearly detected in the F090W image and thus is not a $z$-band dropout source; instead, the galaxy has a photometric redshift solution of $z\sim1.6$ and is best fitted with a star-forming galaxy template ($\logm\sim8.5$), while our IFU cube does not reveal strong emission lines.

We fit each emission line of interest with a Gaussian function. Our basic strategy is to define a wavelength window for each line, model the underlying continuum spectrum (either source origin or residual background) by a polynomial fit, and subtract it from the one-dimensional spectrum before the Gaussian fit is carried out. The model includes amplitude, line width, and redshift parameters. The \oiii$_{\lambda\lambda4959+5007}$ doublet lines are modeled with a fixed line ratio (1:3), with common line width and redshift parameters. The \oii$_{\lambda\lambda3726+3729}$ doublet lines are modeled with common line width, but without their line ratio fixed (see Sec.~\ref{sec:metal}). For emission lines other than those two doublet pairs, we fix the redshift parameter to the value determined by the \oiii-doublets and leave the amplitude and width as free parameters. 

The flux of each line is measured by integrating the modeled Gaussian component, and the flux error is estimated by summing the error weighted by the amplitude of the Gaussian model in quadrature. The uncertainty associated with the modeled continuum is also added. In the following analysis, we adopt flux measurements when the line is detected (signal-to-noise ratio SNR\,$\geq1.5$); for those not detected, we proceed our analysis with the 1.5-$\sigma$ upper limit. We securely detect the \oiii\ doublets in all regions. \hb\ in four and \oii\ in three. In addition, the auroral \oiiif\ is detected in ZD12-W, allowing us to measure the electron temperature (Sec.~\ref{sec:metal}). 

We find that all five regions are within a small distance along the radial direction ($\Delta v = 1500$\,km\,s$^{-1}$). The separation between ZD3 and ZD6 is even smaller, $\delta v = 350$\,km\,s$^{-1}$, confirming that they are interacting, rather than by chance alignment in the sky plane.

The measured line fluxes are corrected for dust attenuation. Although it is ideal to determine the correction on the basis of the ratio of Balmer lines, no more than one line is confidently detected in our cube, except for ZD12-W. We thus apply the correction derived from our resolved SED analysis (Sec.~\ref{sec:sed}), after rescaling it by a factor of 2.27 \citep[][also \citealt{morishita24b}, which found the assumption is valid using $z>3$ galaxies]{calzetti00} that accounts for continuum and emission line dust effects. Dust-corrected line fluxes are reported in Table~\ref{tab:linefluxes}.

\begin{figure}
\centering
	\includegraphics[width=0.48\textwidth]{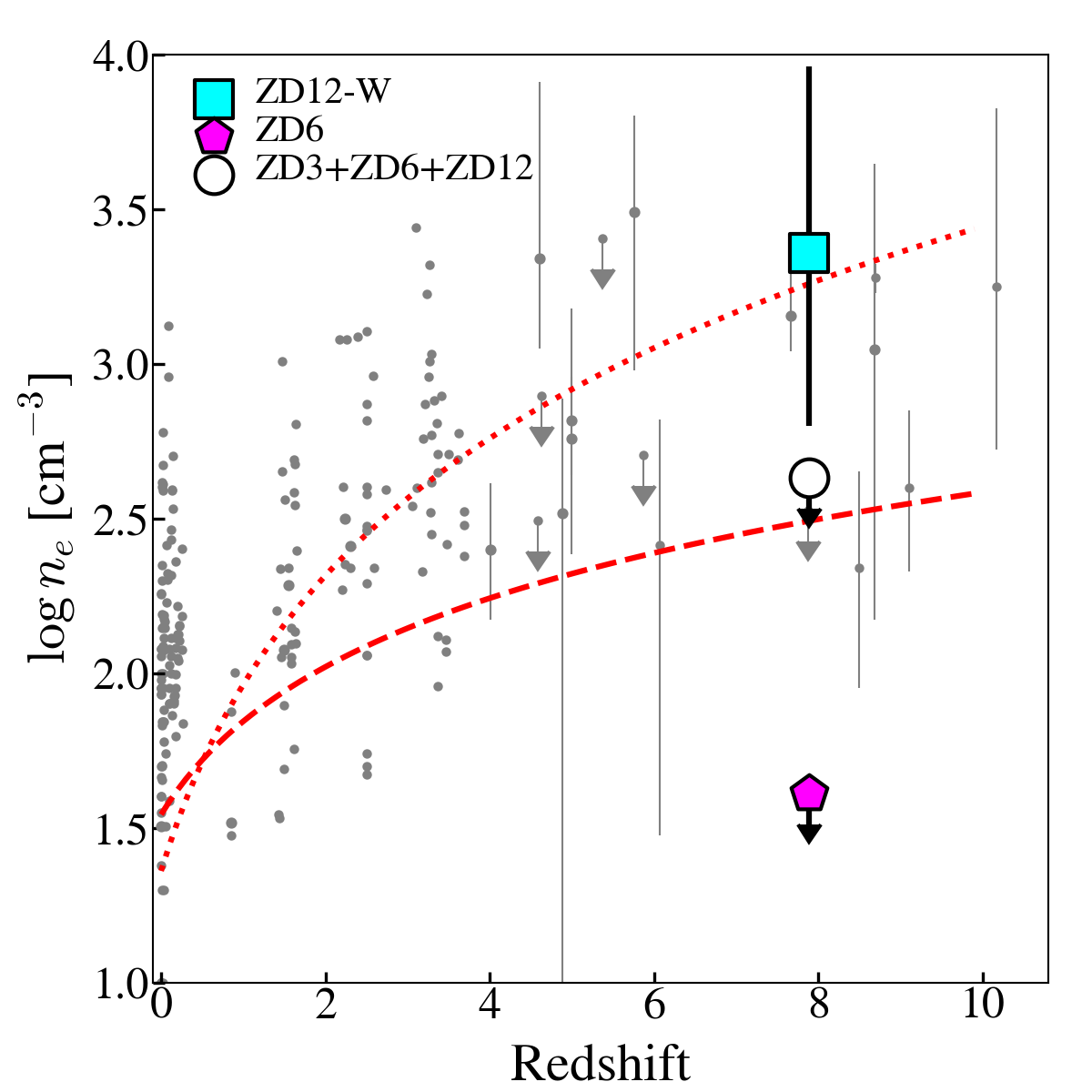}
	\caption{
    Electron density measurements of ZD12-W (cyan square), ZD6 (magenta pentagon), and the total (ZD3+ZD6+ZD12; white circle) shown along with those in the literature (gray dots; data compiled in \citealt{abdurrouf24}). The evolution curves ($\propto(1+z)$ and $\propto(1+z)^2$) derived in \citet{isobe23} are shown.
    }
\label{fig:ne}
\end{figure}

\subsection{Metallicity Measurements} 
\label{sec:metal}

\subsubsection{ZD12-W: Electron Temperature, Density, and Direct Oxygen Abundance} 
\label{sec:Tene}
Our data cube reveals the detection of \oiiif\ in ZD12-W. The auroral line provides a reliable method to derive the electron temperature and from that the oxygen metallicity, known as the $direct$-$T$ method \citep[e.g.,][]{Izotov06}. Therefore, while we rely on a strong-line calibration method for other \oiii-emitting regions, we derive the direct metallicity for ZD12-W for comparison.

The direct method requires knowledge of the electron density $n_e$, which can be measured by the flux ratio of the \oii-doublet lines. The line ratio is measured as $I_{3729}/I_{3726}=0.71\pm0.27$, where $I$ represents the corresponding line intensity. By iteratively solving the equations as in \citet[][also \citealt{osterbrock89}]{Izotov06}, we obtain the electron temperature of O$^{++}$, $T_e=2.4_{-0.4}^{+0.7}\times10^4$\,K, and the electron density $n_e = 2200_{-1600}^{+6700}\,{\rm cm^{-3}}$. The larger error bar of the upper bound density is due to the fact that the adopted equation has a much steeper slope at $n_e\simgt1000\,{\rm cm^{-3}}$ (i.e., $j_{3729}/j_{3726}\simlt0.5$). The measured density is high, but not exceptional among high-$z$ galaxies studied with JWST \citep[e.g.,][]{isobe23,senchyna23,topping24,abdurrouf24,li24}, as shown in Fig.~\ref{fig:ne}. Interestingly, ZD6, the only other source that has a robust doublet measurement, shows a very high line ratio, $I_{3729}/I_{3726}=1.88\pm0.50$. This leads us to $n_e\simlt40\,{\rm cm^{-3}}$ assuming $T_e=10^4$\,K, exhibiting a stark contrast to the one measured for ZD12-W's. Similarly, using the line flux measured in the total (ZD3+ZD6+ZD12) spectrum, we obtain $n_e\simlt430\,{\rm cm^{-3}}$. 

We then estimate the O$^+$ temperature from the O$^{++}$ temperature to $1.5\times10^4$\,K by following \citet{osterbrock89}. The element abundance of O$^{++}$ and O$^{+}$ is derived by using the relation in \citet{Izotov06}, and we obtain \logoh\,$+12=7.41_{-0.17}^{+0.19}$. We have examined the effect of the density uncertainty, by adopting $n_e=1\,{\rm cm^{-3}}$ and $10000\,{\rm cm^{-3}}$, and indeed found that the variability of the result stays within the uncertainties quoted. 

\subsubsection{Oxygen Abundance via Strong Line Calibration Method} 
\label{sec:metal-sl}
For other \oiii-emitting regions that do not detect \oiiif, we use a strong line calibration method to derive metallicity. We use the formula presented in \citet{sanders24}, which offers calibration by using the sample of $z=2$--9 galaxies equipped with direct-T metallicity measurements. 
Using $R3=\log (\oiiir$\,/ \hb), we obtain the metallicity for four regions while ZD6-E is inferred as an upper limit, due to the non-detection of \hb. Using $R23=\log$\,((\oiii$_{\lambda\lambda4959,5007}$ + \oii$_{\lambda\lambda3726,3729}$) / \hb), we obtain similar metallicities, while ZD6-E remains undetermined due to the non-detection of both \hb\ and \oii. The metallicity measurements are reported in Table~\ref{tab:phys}. Using the calibrations presented in \citet{chakraborty24}, we find similar but $\simlt0.1$\,dex higher metallicities for all cases. As such, our analysis and conclusion below are not affected by the choice of calibrators and line indexes.

The indirect metallicity measurement for ZD12-W is $7.5 \pm 0.2$, which is consistent with the one derived by the direct-$T$ method within the error bar. We also measure metallicity for the two integrated systems (ZD12 and ZD3+ZD6, i.e., yellow regions in Fig.~\ref{fig:stamp}). 

\begin{figure}
\centering
	\includegraphics[width=0.48\textwidth]{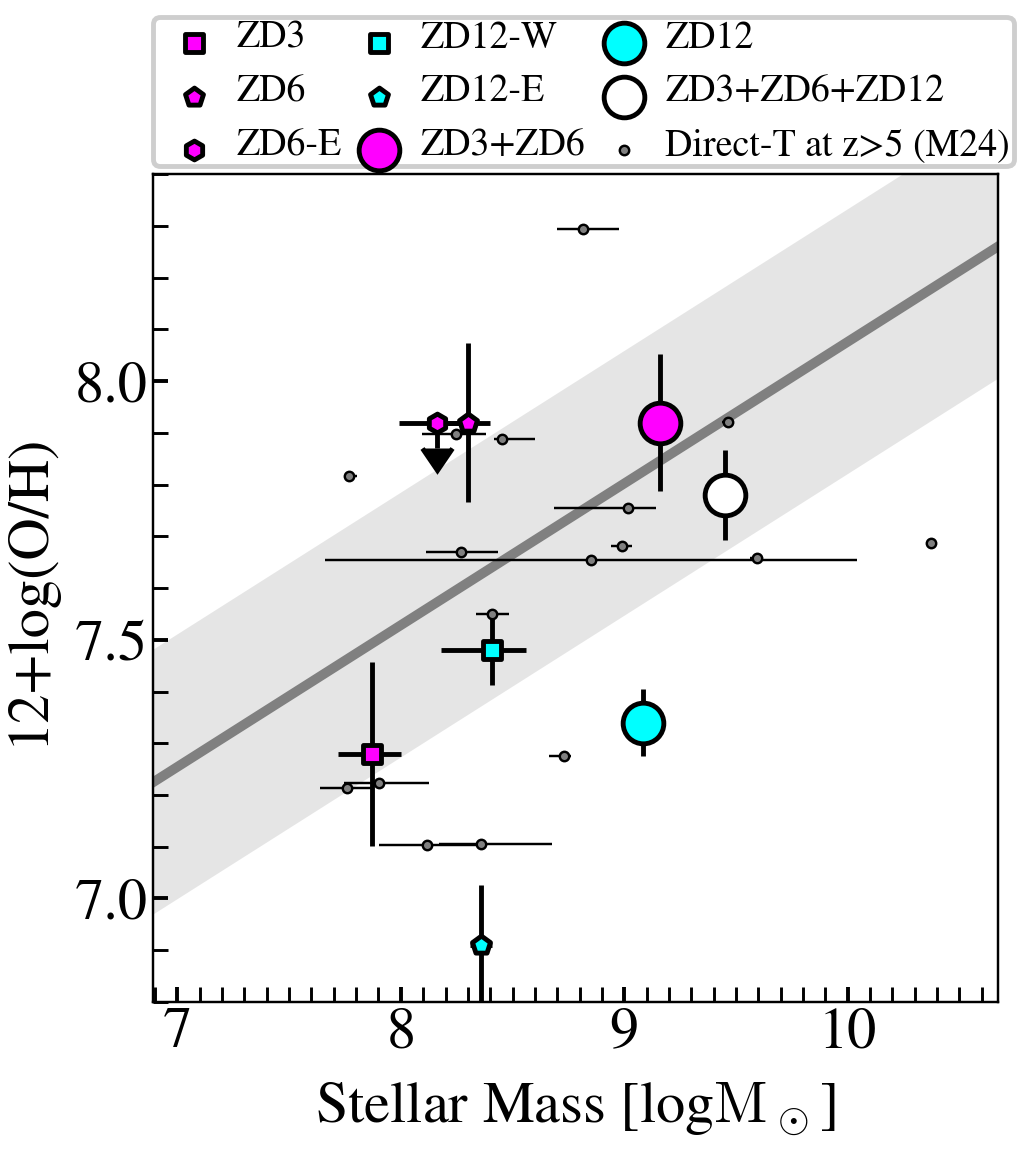}
	\caption{
    Distribution of our sample in the stellar mass--metallicity plane. The mettallicities are measured using the $R3$ calibration presented by  \citet{sanders24}. Large symbols are for measurements over integrated regions and small symbols for individual \oiii-emitting regions (yellow regions and cyan apertures in Fig.~\ref{fig:stamp}, respectively). The mass--metallicity relation \citep{morishita24b} derived for galaxies at $z>5$ (gray dots) is shown. 
    }
\label{fig:MZ}
\end{figure}

The metallicity measurements from the R3 calibration are shown in Fig.~\ref{fig:MZ} by combining with stellar mass measurements obtained for the exact same defined region (see Sec.~\ref{sec:sed}). The two integrated regions are located above and below the mean relation at similar redshifts \citep{morishita24b}, $\sim0.6$\,dex apart from each other. The metallicity measurements of individual \oiii\ regions also show large deviations, $\sim0.6$\,dex within the same galaxy, while the total measurement is represented by the region with higher metallicity. The observed scatters are discussed in Sec.~\ref{sec:disc}.

\subsection{Pixel-by-Pixel SED Analysis}\label{sec:sed}
We infer spectral energy distributions (SEDs) within the detected region of our sample galaxies using the JWST+HST photometric data. Firstly, the spatially resolved segments are defined by the voronoi binning method \citep{cappellari03}, by requiring a minimum SNR of 5 in the F444W image. Fluxes and flux errors in each region are estimated using the PSF-matched images. We use the SED fitting code {\tt gsf} \citep[ver1.9;][]{morishita19}, which allows flexible determinations of the SED by adopting binned star formation histories (often referred to as {\it non-parametric}), and adopt the SMC dust curve \citep{gordon03}. We follow the same procedure as described in \citet{morishita24b}.

Physical quantities of each component, e.g., stellar mass and UV luminosity, are calculated by integrating the derived quantity within the defined region. The UV luminosity is corrected for dust attenuation using the $\beta_{\rm UV}$ slope, which is measured by using the posterior SED, as in \citet{smit16}:
\begin{equation}
    A_{1600} = 4.43 + 1.99\,\beta_{\rm UV}.
\end{equation}
The attenuation corrected UV luminosity is then converted to SFR via the relation in  \citet{kennicutt98}:
\begin{equation}
    {\rm SFR\,[M_\odot\,yr^{-1}]} = 1.4 \times 10^{-28} L_{\rm UV}\,[{\rm erg\,s^{-1}\,Hz^{-1}}].
\end{equation}
Since the relation is essentially for the Salpeter IMF, the estimated SFR is corrected for the Chabrier IMF, by multiplying a factor of 0.63, the factor estimated from {\tt fsps}.

By using the \hb\ flux in Sec.~\ref{linefit} and the UV luminosity derived here, we can measure the ionization photon production efficiency in each \oiii-emitting region \citep[e.g.,][]{schaerer16,Prieto-Lyon23}: 
\begin{equation}
\xi_{\rm ion} = {N_{\rm LyC}/{L_{\rm UV}}}\,\mathrm{[s^{-1}/erg s^{-1} Hz^{-1}]}
\end{equation}
where $N_{\rm LyC}$ is the total ionizing photons of Lyman continuum and $L_{\rm UV}$ intrinsic UV luminosity density measured at rest-frame 1500\,\AA\, both corrected for attenuation. Since the luminosity in the optical hydrogen recombination lines is proportional to the number of Lyman-continuum photons absorbed in the galaxy, the direct measurement of $N_{\rm LyC}$ is proportional to the ionizing photon escape fraction, $f_{\rm esc}$:
\begin{equation}
N_{\rm LyC} = 2.1\times10^{12}(1-f_{\rm esc})^{-1} L({\rm H\beta}),
\end{equation}
where we here adopt $f_{\rm esc}=0$. 

Similarly, using the best-fit SED, we estimate rest-frame equivalent widths (EW) of \oiii-doublet emissions. We note that continuum flux is not detected in the IFU cube in any of the \oiii-emitting regions. This means that our EW measurement is rather {\it indirect} and susceptible to systematics such as absolute flux calibration in the cube. However, we have validated the absolute flux accuracy by comparing the measured line fluxes with the flux excess seen in two broadband photometry, F410M and F444W (where F444W captures the \hb+\oiii\ and F410M does only continuum). The uncertainty in the continuum flux level, estimated in the SED modeling, is also added to the EW measurement uncertainties. The measurements are reported in Table~\ref{tab:phys}.

\begin{figure*}
\centering
    \includegraphics[width=0.95\textwidth]{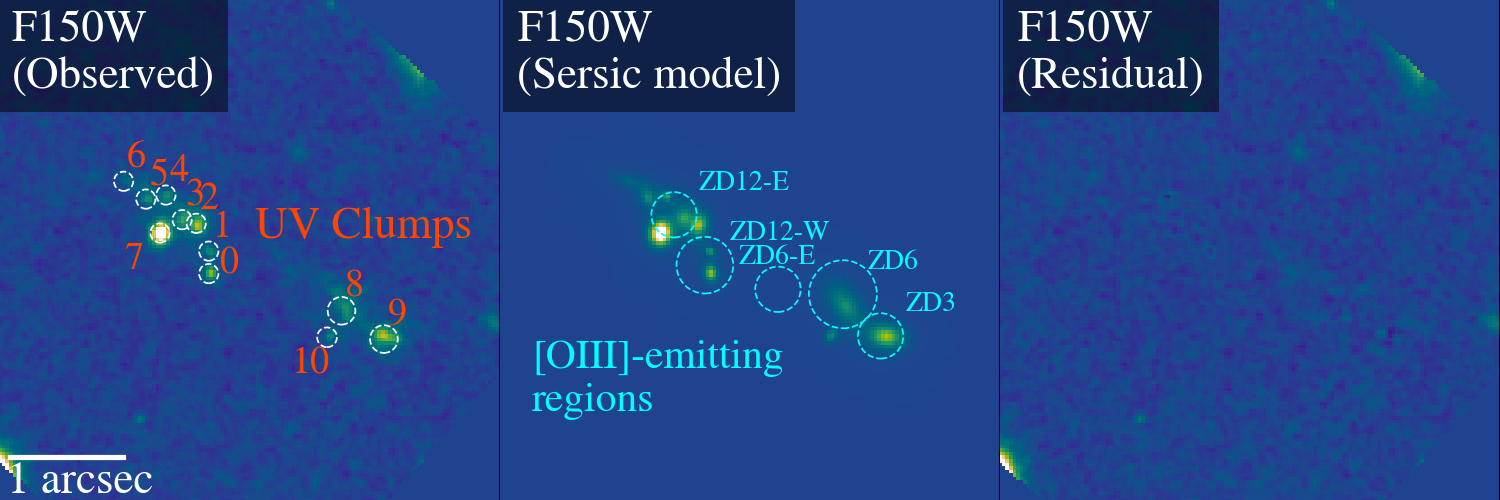}
	\caption{
    Two-dimensional light profile of the system in the NIRCam-F150W band, corresponding to $\sim1700$\,\AA\ rest-frame. \#7 is a foreground galaxy at $z\sim1.6$. ($Left$): observed image. Individual components fitted with {\tt galfit} are identified (circles). ($Middle$): Modeled light profile. The \oiii-emitting regions identified in the IFU data are overlaid (cyan circles) for comparison of the relative positions. ($Right$): Residual image.
    }
\label{fig:galfit}
\end{figure*}

\begin{figure}
\centering
	\includegraphics[width=0.49\textwidth]{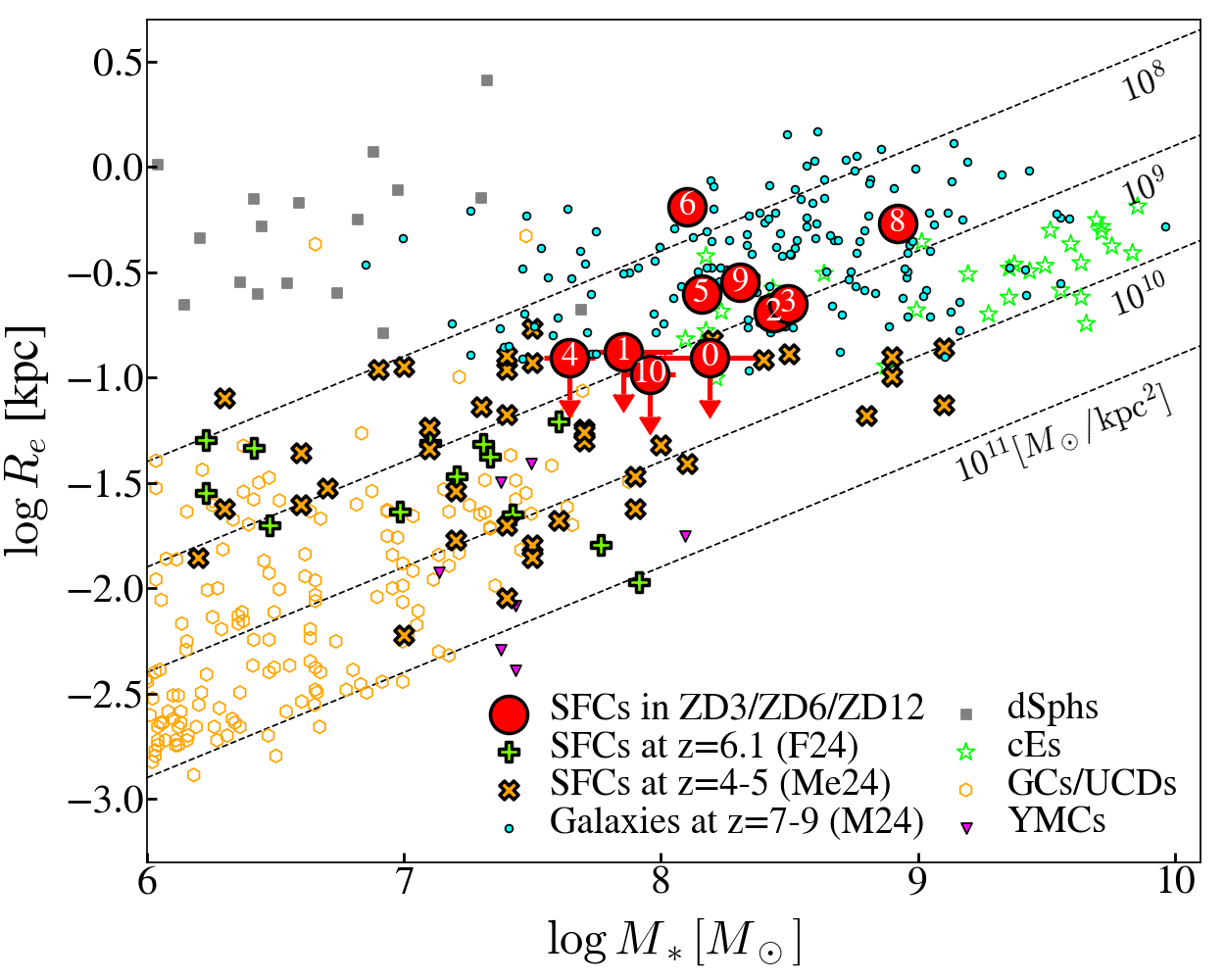}
	\caption{
    Size-mass distributions of UV-bright clumps (red circles). Half-light radii $R_e$ are measured by two-dimensional profile fitting (Fig.~\ref{fig:galfit}). Six UV clumps (\#2, 3, 5, 6, 8, 9) have comparable size with individual galaxies (cyan dots; \citealt{morishita24}) for the given stellar mass. Four (\#0, 1, 4, 10) are unresolved (upper limits), locating near the boundary where local compact ellipticals (lime open stars) and ultra-compact dwarfs/globular clusters (yellow circles) locate (taken from \citealt{norris14}). For comparison, star-forming clusters within a $z=6.1$ galaxy system \citep[][]{fujimoto24} and those within $z=4$--5 galaxies \citep{messa24a} are also shown (cross symbols).
    }
\label{fig:sizemass}
\end{figure}

\section{Discussion}\label{sec:disc}
\subsection{On the Scatter in Metallicity Distribution}\label{sec:mass-metal}

We have seen in Sec.~\ref{sec:metal} that the measured metallicity of the \oiii-emitting regions distributes a wide range, from $\log {\rm O/H}+12<7$ to $\sim8$ (Fig.~\ref{fig:MZ}). Although the sample number here is small, the large scatter observed is qualitatively consistent with previous studies using non-IFU spectra, reporting $\sim0.3$\,dex \citep[e.g.,][]{morishita24b}. \citet[]{morishita24b} attributed it partially to the elevated star-formation activity in early galaxies (see below; also Sec.~\ref{sec:intro}) but also to the aperture effect by the MSA slit. Our IFU observations confirm that large variation of metallicity is present even within in single galaxy systems in these early times. This qualitatively supports the previous interpretation that observed scatters around the MZ relation are partially due to the limited aperture size of MSA.

Notably, the total metallicities measured in the integrated spectra are more represented by enriched components. The integrated ZD3+ZD6 system is at \logoh\,$+12\sim7.9$, whereas ZD3 is characterized with a much lower metallicity, \logoh\,$+12\sim7.3$. Similarly, the integrated ZD12 system is characterized by a higher metallicity, \logoh\,$+12\sim7.3$, than ZD12-E alone, \logoh\,$+12\simlt6.9$. This is a natural consequence of the MZ relation that more massive regions (i.e., more luminous) outshine less enriched (fainter) regions, particularly in young galaxy systems. 
Interestingly, the region with the least enrichment, ZD12-E with an upper limit \logoh\,$+12\simlt7$, is offset for $\sim1$\,kpc from the brightest region of ZD12 in NIRCam imaging. Since MSA masks are configured to source light-weighted positions by default, this presents a challenge in current MSA mask configurations in finding pristine regions of galaxies.

Although this type of observational bias is known for low-$z$ observations in the presence of mild metallicity gradients \citep[e.g.,][]{tremonti04,belfiore17}, the impact is likely more significant at high redshift because of the time available for chemical mixing within ISM. \citet{morishita24b} reported the redshift evolution of scatter amplitude, from $\sim0.1$ at $z\sim0$ \citep{curti20}, to $\sim0.2$ at $\sim0.2$\,dex at $3<z<5$ and $\sim0.3$\,dex at $z>5$ \citep[also][]{curti23}. The impact of star formation feedback, especially in low-mass systems of shallow potential, is therefore expected to be more significant at higher redshifts. \citet[][]{pallottini24} show that high stochasticity in high-$z$ star formation \citep[also][]{shen23,mirocha23,munoz24} enables the system to have large variations of metallicity, making the mass--metallicity relationship less constrained (also see Sec.~\ref{sec:size}).

On the other hand, the metallicity of ZD6 is observed considerably high, \logoh\,$+12\sim8$, or $10\,\%$ solar. This, given the age of the universe $\sim660$\,Myrs, suggests that metal enrichment within some sub-galactic regions could be much more rapid than other regions in the same system. In particular, ZD6 is located near the center of an ongoing merger, where an accelerated gas cycle is expected. Our SED analysis of resolved pixels has revealed an increase in $A_V$, which is another evidence of fast enrichment.

As redshift increases, so does the merger rate \citep[][]{fakhouri2008,Rodriguez-Gomez15}, and thus galaxy assembly is progressively driven by mergers rather than smooth accretion \citep[see also][for recent observations]{hsiao23,boyett24,dalmasso24,sugahara24}. As a result, the chemical enrichment of the system is significantly affected by individual merger events (and also the properties of independent building blocks), rather than by smooth accretion of material and self-recycling, increasing the metallicity variation. As such, in addition to the general increase in star formation intensity in the early universe, our study showcases that an increased fraction of merger-driven evolution would also contribute to the observed scatter in the mass--metallicity relation. A large data set of IFU observations may shed light on the relative contribution of each evolutionary channel.

\subsection{Compact UV-bright Clumps as the Origin of Strong \oiii\ Emissions}\label{sec:size}

NIRSpec IFU is limited to the spatial resolution of $\sim0.\!''15$ ($\sim370$\,pc with $\mu\sim2$ magnification), leaving the nature of the \oiii-emitting regions rather undetermined. In the NIRCam images (Fig.~\ref{fig:stamp}), which offer $\sim5\times$ higher resolution, we see multiple UV-bright clumps in both integrated ZD12 and ZD3+ZD6 systems, at a smaller scale than the \oiii\ regions. 

To study in detail these UV-bright clumps, we fit the two-dimensional light distribution in the F150W image, which corresponds to $\sim1700$\,\AA\ rest-frame, by using {\tt galfit} \citep{peng02,peng10}. Each light profile of 10 UV clumps (and one foreground galaxy) is well fitted by a single-S\'ersic component (Fig.~\ref{fig:galfit}). Four (\# 0, 1, 4, 10) have effective radii smaller than the PSF size ($\simlt2$\,pixels, or $R_e\simlt0.\!''06$). We consider those clumps unresolved and quote the PSF's FHWM as their upper limit sizes. 

In Fig.~\ref{fig:sizemass}, we show the size distribution of the UV clumps as a function of the stellar mass. The stellar mass of each clump is estimated by using the mass-to-light ratio derived in Sec.~\ref{sec:sed} and scaling it with the total flux modeled with {\tt galfit}. The four unresolved UV clumps are found in the regime where relatively massive clusters in the local universe are located \citep{norris14}. However, with the spatial resolution available here, the detailed classification of those unresolved clumps is undetermined. The other six UV clumps (\# 2, 3, 5, 6, 8, 9) have a size comparable to that of individual galaxies \citep{morishita24}. 

\begin{figure*}
\centering
    \includegraphics[width=0.49\textwidth]{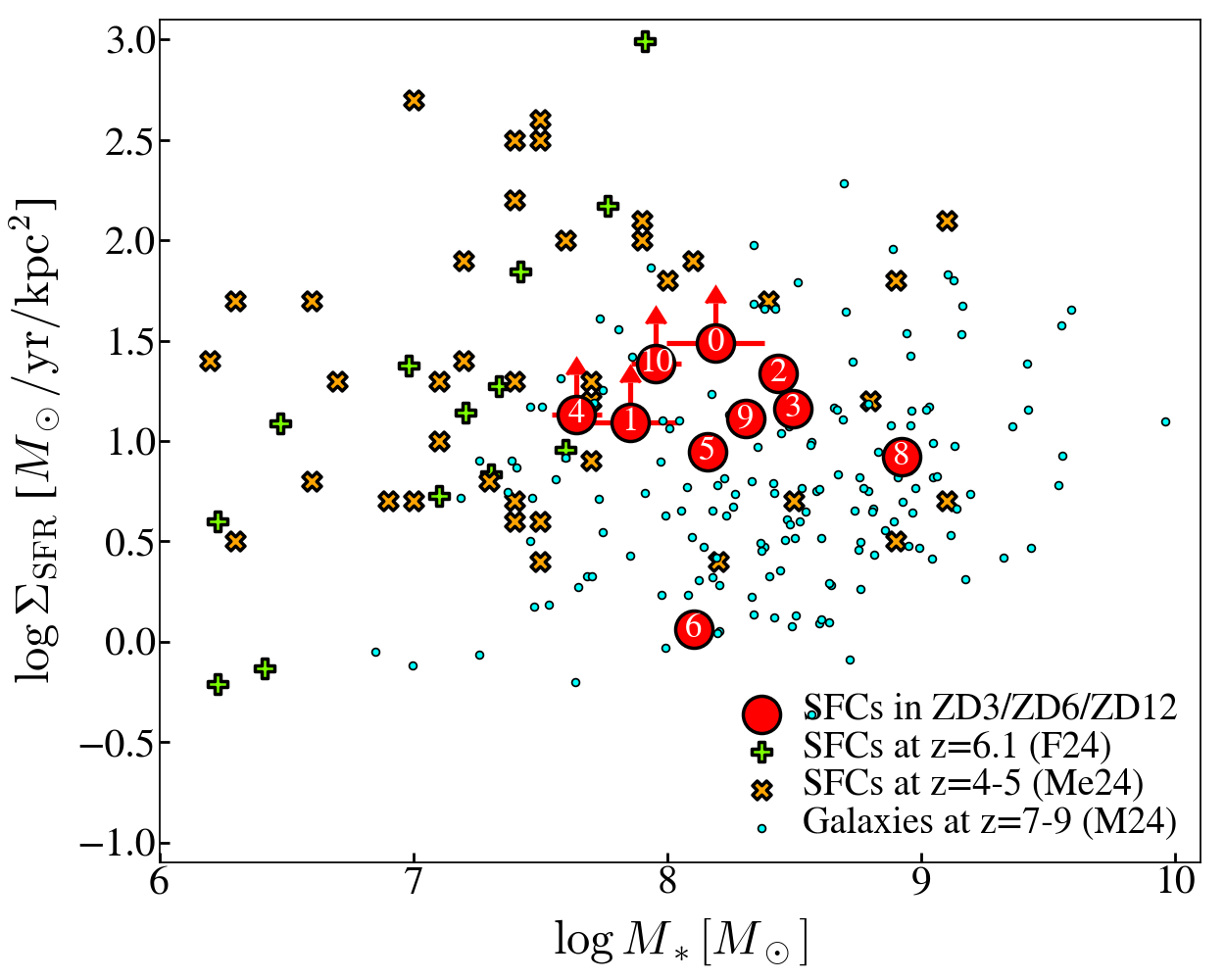}
    \includegraphics[width=0.49\textwidth]{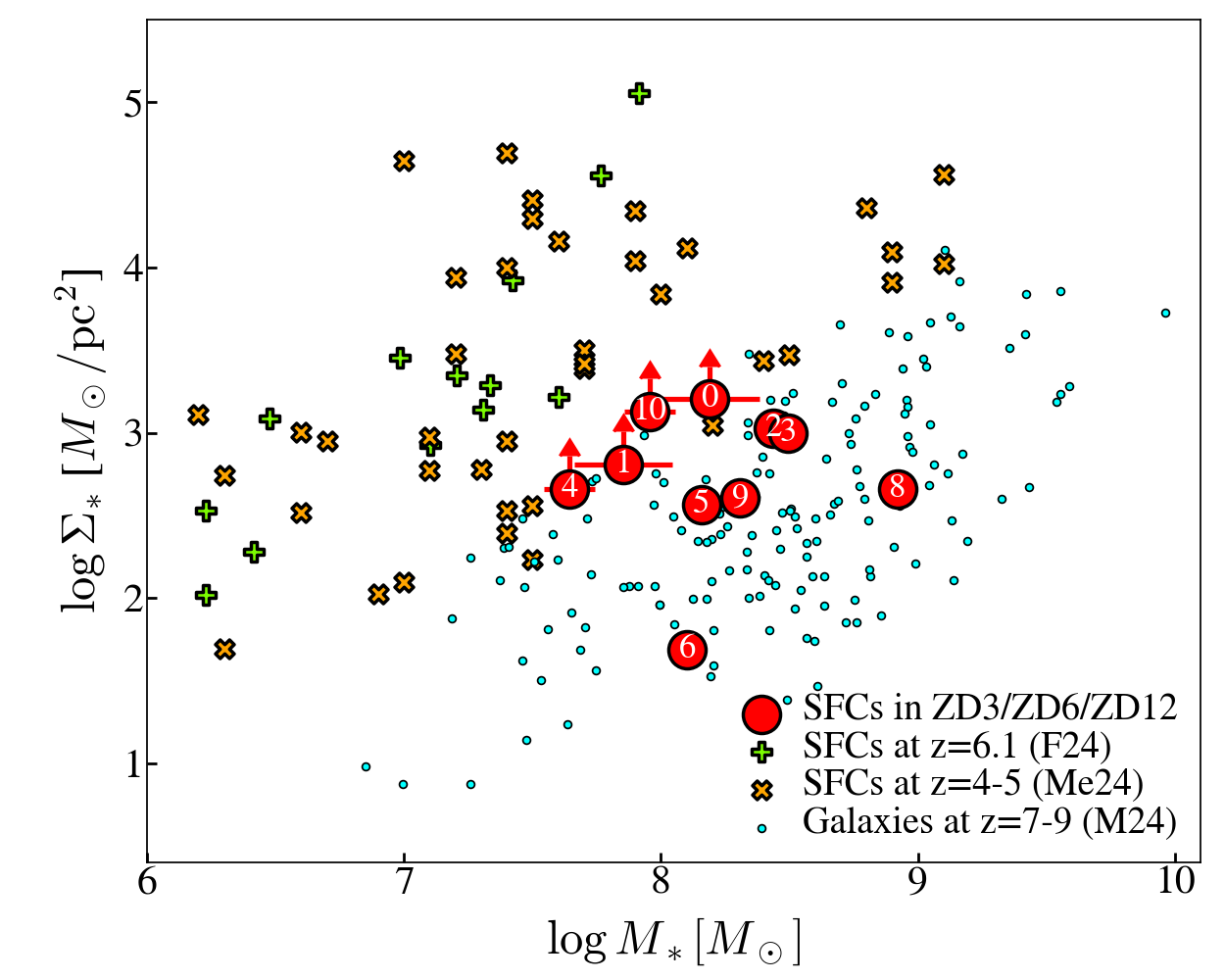}
	\caption{($Left$): Star formation surface density as a function of stellar mass. Symbols are the same as in Fig.~\ref{fig:sizemass}. ($Right$): Same as the left panel but for stellar mass surface density. 
    }
\label{fig:sf}
\end{figure*}

Those unresolved clumps are characterized with a high star formation surface density $\Sigma_{\rm SFR}$ ($>10$--$100\,M_\odot/{\rm yr/kpc^2}$) and a surface mass surface density $\Sigma_{\rm *}$ ($>500$--$1600\,M_\odot/{\rm pc^2}$; Fig.~\ref{fig:sf}). The observed densities are on the upper end of individual galaxy populations at $z>5$ \citep[e.g.,][]{morishita24}, even reaching the realm of star clusters. 

Recent observations have seen some evidence of star clusters forming in the early environment through gravitational lensing \citep[e.g.,][]{vanzella22,vanzella23a,mevstric22,adamo24,messa24a,fujimoto24}. For example, \citet{messa24} recently presented NIRSpec/IFU observations and revealed star clusters of $R_e\sim3$--10\,pc within a strongly lensed galaxy system at $z=6.14$. These clusters are strongly line-emitting and are characterized by blue UV slopes, $\beta_{\rm UV} \simlt -2.5$ and $\simlt0.1\,Z_\odot$. 

In this context, the nature of star formation in our unresolved clumps is of particular interest. The UV clump \#0 and 1 are both located within the \oiii-emitting region (ZD12-W), one of the metal-poor regions (\logoh\,$+12=7.4$ from the direct-T method) with high \oiii-equivalent width, EW$_0=670\pm60$\,\AA. Remarkably, the measured ionizing photon production efficiency (Sec.~\ref{sec:sed}) for ZD12-W is considerably high with $\log \xi_{\rm ion} = 25.7$\,Hz/erg for its $M_{\rm UV}$ when compared to those of the literature \citep[e.g.,][]{vanzella24}. 
Using the linear relation between $\xi_{\rm ion}$ and $M_{\rm UV}$ derived by \citet[][]{Prieto-Lyon23}, a rather moderate efficiency, $\sim25.4$, would be expected for ZD12-W's $M_{\rm UV}$. This suggests a higher escape fraction in ZD12-W than assumed here, i.e., $f_{\rm esc}\sim0.5$ would be required to meet the expected $\xi_{\rm ion}$. Alternatively, if the observed \hb\ originates mainly in one of the two UV clumps (which have $M_{\rm UV}\sim -17.7$ and $-16.9$\,mag), it would make the deviation smaller, but still not completely gone.

Star formation in such dense environments plays a crucial role in chemical enrichment and may be responsible for the diverse properties observed among our samples. Recent JWST studies have also illuminated the origin of star clusters by identifying Nitrogen-rich ISM in the early universe \citep[e.g.,][]{topping24,stiavelli24}. A detailed study of the chemical compositions within those UV clumps would thus give us further insight into the nature, e.g., whether they are giant star-forming clumps (similar to those at lower redshifts) or even smaller-scale populations such as star clusters.

Theoretical studies investigated the nature of star-forming clumps seen at high redshifts in relation with metallicity. \citet[][]{sugimura24} show that high gas density can be reached within an environment of metal-poor star formation because of the intense far-UV radiation field therein. This is qualitatively consistent with the nature of ZD12-W --- low metallicity, high electron density, and high surface densities. \citet[][]{nakazato24} show that short-lived, UV-bright clumps can form through merger, which may also be relevant to the case of the ZD3+ZD6+ZD12 system. Similarly, \citet[][]{marconcini24} investigated the moderate anti-correlation between $\Sigma_{\rm SFR}$ and \logoh\ seen in a galaxy system CR7. These authors attributed it to the inflow of pristine gas, which would lower the metallicity and trigger intense star formation, though the system's global metallicity is much more enriched (\logoh\,$+12\sim8$) than our case here. Different mechanisms responsible for high $\Sigma_{\rm SFR}$ may be in action at different metallicity regimes. 


\subsection{Revisiting \id\ as a protocluster}\label{sec:pc}

Lastly, we revisit the characterization of \id\ as a protocluster system. \id\ was reported in \citet{morishita23b} for the spectroscopic confirmation of seven member galaxies at $z=7.88$ \citep[also][]{roberts-borsani22}. The following studies have identified two \citep[YD1, s1;][]{hashimoto23} and one \citep[ZD4;][]{chen24} additional members. With the new addition of ZD12 in this study, now 11 member galaxies are confirmed (Fig.~\ref{fig:stamp}), distributed within a small projected region ($r\sim60$\,pkpc, in the source plane) and a redshift slice ($\delta_z=0.019$, or $\sim25$\,pMpc). Following \citet[][]{morishita23b}, we here aim to estimate the overdensity factor, the total halo mass, and the system velocity dispersion.

The overdensity factor, defined as $\delta=(n-\bar{n})/\bar{n}$, represents the excess of the surface number density from the field average. We use $R=60$\,pkpc, which defines the extent of the member galaxies in the source plane, and the redshift slice of $\delta_z=0.019$. One of the member galaxies, s1, is excluded from the calculation, as this source is rather considered as a sub-region and does not qualify for ``galaxy'' count. For field reference, we use the luminosity function at $z\sim8$ presented in \citet{bouwens21} and integrate it down to $M_{\rm UV}\sim-17.4$\,mag, the lowest magnitude among the confirmed member galaxies (except for s1). For the area within $R=60$\,pkpc, we expect $\bar{n}=0.2_{-0.4}^{+0.5}$ galaxies in average fields. With the reference number for the average field, we find $\delta = 44_{-31}^{+89}$ for \id, where the uncertainties quoted represent Poisson noise.

To estimate the total halo mass of \id, we estimate the halo mass of the individual components from the halo-mass galaxy-luminosity relation \citep{mason23}. Using the relation and adding the halo mass estimates of all confirmed members, we obtain the total halo mass estimate $5.8_{-0.3}^{+0.2}\times10^{11}\,M_\odot$, but this is likely a lower limit given that there are candidate selected photometrically that are not spectroscopically followed.

In addition, we can take advantage of the spectroscopic data to obtain an estimate of the velocity dispersion for the whole system. We adopt a simple Gaussian estimator and bootstrap method to derive the uncertainty \citep{Beers90}, obtaining $1100\pm500\,\kms$, consistent with the original estimate in \citet{morishita23b}. We caution the reader that the system is likely not virialized, and that in computing this quantity we are assuming the spread in redshift with respect to the mean is due to motion as opposed to distance along the line of sight. 

We note that the only \ly-emitting member galaxy, ZD4 \citep[][]{chen24}, is located in an off-core region of the protocluster (Fig.~\ref{fig:stamp}). 

\section{Summary}\label{sec:sum}
In this study, we reported our JWST/NIRSpec IFU observations on a merging galaxy system in the core of \id\ at $z=7.88$, the most distance protocluster to date. The IFU cube revealed five \oiii-emitting regions, including those from a newly confirmed protocluster member ZD12. We found large scatters in the measured metallicity in those emitting regions, for $\sim1$\,dex. We discussed the origin of the large scatter and attributed it to a fast gas cycle from intense star formation and merger-driven growth. The presence of large variations in metallicity, even within a single galaxy ($\sim0.6$\,dex), posed a challenge to current slit spectroscopy with NIRSpec MSA, in finding metal-poor star formation in the early universe \citep[e.g. those found in][]{vanzella23}, and also in obtaining the {\it true} metallicity of a given galaxy system, due to the limited aperture size of the MSA.

We have seen that ZD12 consists of multiple UV-bright clumps, revealed in the high-resolution NIRCam F150W image. Four of them are unresolved and characterized with high masses or star-formation surface densities. Two unresolved UV clumps are likely responsible for the strong \oiii\ in ZD12-W, which is characterized by a metal-poor ISM (\logoh\,$+12=7.4$), high electron density ($n_e=2200_{-1600}^{+6700}$), large \oiii-equivalent width (EW$_0=670\pm60$\,\AA), and high ionizing photon efficiency ($\log\xi_{\rm ion}=25.7$\,Hz/erg). The nature of those compact star-forming clumps remains undetermined. Detailed chemical investigations that involve other element species, such as Nitrogen and Argon would be required \citep[e.g.,][]{stiavelli24}.

Lastly, we revisited the characterization of \id, as a large-scale structure. Its overdensity factor, halo mass, and velocity dispersion have been updated by including the newly identified ZD12 and other new member galaxies in the literature; however, we note that our characterization was made with the measurements within the very central core of the protocluster, and the membership determination is still incomplete, because of the nature of the existing observations. Extending the search of member galaxies from further distances and in different wavelengths (e.g., submillimeters for heavily obscured sources) will provide us with a more comprehensive view of the protocluster.

\begin{deluxetable*}{ccccccc}
\tablecaption{
Fluxes are in units of $10^{-20}$\,erg/s/cm$^2$. Flux errors are $1\,\sigma$.
}
\tablehead{
\colhead{ID} & \colhead{${\rm [OII]_{3726}}$} & \colhead{${\rm [OII]_{3729}}$} & \colhead{${\rm H\gamma_{4340}}$} & \colhead{${\rm [OIII]_{4363}}$} & \colhead{${\rm H\beta_{4861}}$} & \colhead{${\rm [OIII]_{4959,5007}}$}
}
\startdata
ZD3 & {$2.9 \pm 1.7$} & {$4.3 \pm 1.8$} & {$1.6 \pm 1.0$} & {$<1.8$} & {$6.5 \pm 2.4$} & {$50.4 \pm 9.0$}\\
ZD6 & {$40.2 \pm 10.4$} & {$75.5 \pm 11.5$} & {$18.2 \pm 7.0$} & {$<13.3$} & {$22.8 \pm 7.7$} & {$382.8 \pm 32.6$}\\
ZD6-E & {$<1.7$} & {$<1.8$} & {$<1.3$} & {$<1.2$} & {$<1.1$} & {$13.3 \pm 4.4$}\\
ZD12-W & {$20.7 \pm 4.3$} & {$14.7 \pm 4.7$} & {$14.0 \pm 3.1$} & {$10.4 \pm 3.0$} & {$40.5 \pm 5.5$} & {$313.9 \pm 21.0$}\\
ZD12-E & {$<1.3$} & {$1.8 \pm 0.8$} & {$<0.7$} & {$<0.9$} & {$6.9 \pm 1.3$} & {$26.1 \pm 4.3$}\\
ZD3+ZD6 & {$44.0 \pm 10.8$} & {$72.6 \pm 11.8$} & {$24.5 \pm 9.2$} & {$15.8 \pm 9.8$} & {$26.8 \pm 7.8$} & {$456.4 \pm 36.4$}\\
ZD12 & {$20.1 \pm 6.2$} & {$17.0 \pm 5.8$} & {$11.7 \pm 2.9$} & {$15.4 \pm 4.6$} & {$60.9 \pm 7.8$} & {$429.4 \pm 29.6$}\\
ZD3+ZD6+ZD12 & {$53.3 \pm 14.1$} & {$74.6 \pm 12.8$} & {$27.6 \pm 11.2$} & {$33.1 \pm 11.1$} & {$60.9 \pm 12.6$} & {$706.6 \pm 51.4$}\\
\enddata
\tablecomments{
Fluxes are in units of $10^{-20}$\,erg/s/cm$^2$. Flux errors are $1\,\sigma$.
1.5-$\sigma$ upper limits are quoted for non-detected lines (SN\,$<1.5$). Lines without the spectroscopic data coverage are marked with ``--". Flux values have been corrected for dust attenuation.
}\label{tab:linefluxes}
\end{deluxetable*}
\startlongtable
\begin{deluxetable*}{cccccccccccc}
\tablecolumns{12}
\tabcolsep=0.06cm
\tablecaption{
Physical properties of \oiii-emitting regions.}
\tablehead{
\colhead{ID} & \colhead{R.A.} & \colhead{Decl.} & \colhead{$z$} & \colhead{$\mu$} & \colhead{$M_\mathrm{UV}$} & \colhead{$\log M_*$} & \colhead{$\log {\rm SFR_{UV}}$} & \colhead{$\log {\rm (O/H)_{R3}}$} & \colhead{$\log {\rm (O/H)_{R23}}$} & \colhead{$\log \xi_{\rm ion}$} & \colhead{EW$_0({\rm [OIII]})$}\\
\colhead{} & \colhead{deg.} & \colhead{deg.} & \colhead{} & \colhead{} & \colhead{mag} & \colhead{$M_\odot$} & \colhead{$M_\odot {\rm yr^{-1}}$} & \colhead{$+12$} & \colhead{$+12$} & \colhead{${\rm Hz\,erg^{-1}}$} & \colhead{${\rm \AA}$}
}
\startdata
ZD3 & 3.6064713 & -30.3810350 & $7.8808$ & $1.96_{-0.06}^{+0.07}$ & $-18.5_{-0.1}^{+0.1}$ &  $7.9_{-0.1}^{+0.1}$ & $0.4_{-0.1}^{+0.1}$ & $7.28\pm0.38$ & $7.33\pm0.51$ & $25.3_{-0.2}^{+0.1}$ & $167_{-31}^{+31}$\\
ZD6 & 3.6065784 & -30.3809330 & $7.8797$ & $1.96_{-0.06}^{+0.07}$ & $-19.5_{-0.1}^{+0.1}$ &  $8.3_{-0.1}^{+0.1}$ & $0.9_{-0.1}^{+0.1}$ & $7.92\pm0.01$ & $8.04\pm0.01$ & $25.2_{-0.2}^{+0.1}$ & $205_{-21}^{+20}$\\
ZD6-E & 3.6067629 & -30.3809210 & $7.8802$ & $1.96_{-0.06}^{+0.07}$ & $-17.5_{-0.1}^{+0.1}$ &  $8.2_{-0.2}^{+0.1}$ & $0.0_{-0.2}^{+0.2}$ & $<7.92$ & --$^\ddagger$ & $<24.7$ & $48_{-15}^{+15}$\\
ZD12-W & 3.6069704 & -30.3808620 & $7.8759$ & $1.96_{-0.06}^{+0.07}$ & $-19.2_{-0.1}^{+0.1}$ &  $8.4_{-0.2}^{+0.1}$ & $0.7_{-0.1}^{+0.1}$ & $7.48\pm0.18$$^\dagger$ & $7.53\pm0.21$$^\dagger$ & $25.7_{-0.1}^{+0.1}$ & $669_{-62}^{+59}$\\
ZD12-E & 3.6070579 & -30.3807390 & $7.8782$ & $1.96_{-0.06}^{+0.07}$ & $-18.1_{-0.1}^{+0.1}$ &  $8.4_{-0.1}^{+0.1}$ & $0.3_{-0.1}^{+0.1}$ & $6.91\pm0.13$ & $<6.83$ & $25.4_{-0.1}^{+0.1}$ & $37_{-6}^{+6}$\\
ZD3+ZD6 & 3.6065249 & -30.3808000 & $7.8799$ & $1.95_{-0.06}^{+0.07}$ & $-19.9_{-0.1}^{+0.1}$ &  $9.2_{-0.1}^{+0.1}$ & $1.5_{-0.1}^{+0.1}$ & $7.92\pm0.01$ & $8.04\pm0.01$ & $24.5_{-0.1}^{+0.1}$ & $99_{-10}^{+9}$\\
ZD12 & 3.6065249 & -30.3808000 & $7.8762$ & $1.95_{-0.06}^{+0.07}$ & $-20.0_{-0.1}^{+0.1}$ &  $9.1_{-0.1}^{+0.1}$ & $1.4_{-0.1}^{+0.1}$ & $7.34\pm0.13$ & $7.34\pm0.14$ & $24.8_{-0.1}^{+0.1}$ & $61_{-6}^{+5}$\\
ZD3+ZD6+ZD12 & 3.6065249 & -30.3808000 & $7.8784$ & $1.95_{-0.06}^{+0.07}$ & $-20.8_{-0.1}^{+0.1}$ &  $9.5_{-0.1}^{+0.1}$ & $1.7_{-0.1}^{+0.1}$ & $7.78\pm0.23$ & $8.04\pm0.24$ & $24.5_{-0.1}^{+0.1}$ & $58_{-6}^{+5}$\\
\enddata
\tablecomments{
Measurements are corrected for magnification (Sec.~\ref{sec:data}).
$\dagger$: Metallicity through the direct-T method is \logoh\,$+12=7.41_{-0.17}^{+0.19}$.
$\ddagger$: \oii\, is not detected, thus the derived metallicity via $R23$ is considered as an upper limit.
}\label{tab:phys}
\end{deluxetable*}
\startlongtable
\begin{deluxetable*}{cccccccccccc}
\tablecolumns{12}
\tabcolsep=0.08cm
\tablecaption{Physical properties of UV clumps.
}
\tablehead{
\colhead{ID} & \colhead{R.A.} & \colhead{Decl.} & \colhead{$\mu$} & \colhead{$m_\mathrm{F150W}$} & \colhead{$M_\mathrm{UV}$} & \colhead{$\log M_*$} & \colhead{$\log {\rm SFR_{UV}}$} & \colhead{$\log R_e$} & \colhead{$n_\mathrm{Sersic}$} & \colhead{$\log \Sigma_{\rm SFR}$} & \colhead{$\log \Sigma_{\rm *}$}\\
\colhead{} & \colhead{deg.} & \colhead{deg.} & \colhead{} & \colhead{mag} & \colhead{mag} & \colhead{$M_\odot$} & \colhead{$M_\odot {\rm yr^{-1}}$} & \colhead{kpc} & \colhead{} & \colhead{$M_\odot {\rm yr^{-1} kpc^{-2}}$} & \colhead{$M_\odot {\rm pc^{-2}}$}
}
\startdata
0 & 3.60694 & -30.38082 & $1.96$ & $28.3\pm0.0$ & $-17.7_{-0.1}^{+0.1}$ & $8.2_{-0.2}^{+0.2}$ & $0.5_{-0.1}^{+0.1}$ & $<-0.9$ & $0.8\pm0.1$ & $>1.5$ & $>3.2$\\
1 & 3.60694 & -30.38077 & $1.96$ & $29.1\pm0.2$ & $-16.9_{-0.1}^{+0.1}$ & $7.9_{-0.2}^{+0.2}$ & $0.1_{-0.1}^{+0.1}$ & $<-0.9$ & $4.0\pm2.7$ & $>1.1$ & $>2.8$\\
2 & 3.60697 & -30.38070 & $1.96$ & $27.5\pm0.0$ & $-18.4_{-0.1}^{+0.1}$ & $8.4_{-0.1}^{+0.1}$ & $0.8_{-0.1}^{+0.1}$ & $-0.7\pm0.4$ & $0.8\pm0.0$ & $1.3\pm0.4$ & $3.0\pm0.4$\\
3 & 3.60701 & -30.38069 & $1.96$ & $27.9\pm0.1$ & $-18.1_{-0.0}^{+0.0}$ & $8.5_{-0.1}^{+0.1}$ & $0.7_{-0.1}^{+0.1}$ & $-0.7\pm0.5$ & $0.8\pm0.1$ & $1.2\pm0.5$ & $3.0\pm0.5$\\
4 & 3.60706 & -30.38063 & $1.95$ & $28.8\pm0.1$ & $-17.2_{-0.0}^{+0.0}$ & $7.6_{-0.1}^{+0.1}$ & $0.1_{-0.1}^{+0.1}$ & $<-0.9$ & $0.5\pm0.1$ & $>1.1$ & $>2.7$\\
5 & 3.60712 & -30.38064 & $1.95$ & $28.3\pm0.2$ & $-17.6_{-0.1}^{+0.1}$ & $8.2_{-0.2}^{+0.2}$ & $0.5_{-0.1}^{+0.1}$ & $-0.6\pm0.6$ & $1.4\pm0.3$ & $0.9\pm0.6$ & $2.6\pm0.6$\\
6 & 3.60718 & -30.38060 & $1.95$ & $28.5\pm0.2$ & $-17.5_{-0.1}^{+0.1}$ & $8.1_{-0.2}^{+0.2}$ & $0.5_{-0.1}^{+0.1}$ & $-0.2\pm1.0$ & $1.7\pm0.5$ & $0.1\pm1.0$ & $1.7\pm1.0$\\
8 & 3.60656 & -30.38091 & $1.96$ & $26.9\pm0.0$ & $-18.9_{-0.1}^{+0.1}$ & $8.9_{-0.1}^{+0.1}$ & $1.2_{-0.1}^{+0.1}$ & $-0.3\pm0.8$ & $1.0\pm0.0$ & $0.9\pm0.8$ & $2.7\pm0.8$\\
9 & 3.60644 & -30.38098 & $1.97$ & $27.0\pm0.0$ & $-19.0_{-0.1}^{+0.1}$ & $8.3_{-0.2}^{+0.1}$ & $0.8_{-0.1}^{+0.1}$ & $-0.5\pm0.6$ & $0.7\pm0.0$ & $1.1\pm0.6$ & $2.6\pm0.6$\\
10 & 3.60660 & -30.38098 & $1.97$ & $29.3\pm0.0$ & $-16.5_{-0.1}^{+0.1}$ & $8.0_{-0.1}^{+0.1}$ & $0.2_{-0.1}^{+0.1}$ & $<-1.0$ & $0.6\pm0.3$ & $>1.4$ & $>3.1$\\
\enddata
\tablecomments{
UV clumps are defined in the NIRCam F150W image (Fig.~\ref{fig:galfit}). 
Measurements are corrected for magnification (Sec.~\ref{sec:data}).
}\label{tab:uvclumps}
\end{deluxetable*}

\section*{Acknowledgements}
We thank Abdurro'uf for sharing the compiled data catalog of electron density measurements in the literature.
Some/all of the data presented in this paper were obtained from the Mikulski Archive for Space Telescopes (MAST) at the Space Telescope Science Institute. The specific observations analyzed can be accessed via \dataset[10.17909/q8cd-2q22]{https://doi.org/10.17909/q8cd-2q22}. We acknowledge support for this work under NASA grant 80NSSC22K1294. Support for this work was provided by NASA through the NASA Hubble Fellowship grant HST-HF2-51492 awarded to AJS by the Space Telescope Science Institute (STScI), which is operated by the Association of Universities for Research in Astronomy, Inc., for NASA, under contract NAS5-26555. AJS also received support from NASA through the STScI grants HST-GO-16773 and JWST-GO-2974.

{
{\it Software:} 
Astropy \citep{astropy13,astropy18,astropy22}, 
gsf \citep{morishita19}, lacosmic \citep{vandokkum01,bradley23},
numpy \citep{numpy}, python-fsps \citep{foreman14}, JWST pipeline \citep{jwst},
SExtractor \citep{bertin96}.
}


\bibliography{output}{}
\bibliographystyle{aasjournal}



\end{document}